\documentclass[preprint, amsmath, amssymb, preprintnumbers, showpacs, showkeys, aps, prl]{revtex4-1}

\usepackage[version=3]{mhchem} 
\usepackage{graphicx}
\usepackage{float}
\usepackage{color}
\usepackage{natmove}

\begin{document}
\preprint{}

\title{3D Dirac semimetals: current materials, design principles and predictions of new materials}
\author{Q.D.Gibson$^1$}
\author{L.M.Schoop$^1$}
\author{L.Muechler$^1$}
\author{L.S.Xie$^1$}
\author{M.Hirschberger$^2$}
\author{N.P.Ong$^2$}
\author{R.Car$^1$}
\author{R.J.Cava$^1$}

\affiliation{$^1$ Princeton University, Department of Chemistry, Princeton, NJ, 08544}
\affiliation{$^2$ Princeton, University, Department of Physics, Princeton, NJ, 08544}

\date{\today}

\begin{abstract}

Design principles and novel predictions of new 3D Dirac semimetals are presented, along with the context of currently known materials. Current materials include those based on a topological to trivial phase transition, such as in TlBiSe$_{2-x}$S$_x$ and Hg$_{1-x}$Cd$_x$Te, Bi$_{1-x}$Sb$_x$, Bi$_{2-x}$In$_x$Se$_3$, and Pb$_{1-x}$Sn$_x$Se. Some more recently revealed materials, Na$_3$Bi and Cd$_3$As$_2$, are 3D Dirac semimetals in their native composition. The different design principles presented each yield novel predictions for new candidates. For Case I, 3D Dirac semimetals based on charge balanced compounds, BaAgBi, SrAgBi, YbAuSb, PtBi$_2$ and SrSn$_2$As$_2$ are identified as  candidates. For Case II, 3D Dirac semi-metals in analogy to graphene, BaGa$_2$ is identified as a candidate, and BaPt and Li$_2$Pt are discussed. For Case III, 3D Dirac semi-metals based on glide planes and screw axes, TlMo$_3$Te$_3$ and the AMo$_3$X$_3$ family in general (A=K, Na, In, Tl, X=Se,Te) as well as the Group IVb trihalides such as HfI$_3$ are identified as candidates. Finally we discuss conventional intermetallic compounds with Dirac cones, and identify Cr$_2$B as a potentially interesting material.
\end{abstract}

\maketitle

\section{Introduction}

Recent advances in condensed matter physics have expanded the usual set of bulk electronic materials beyond metals, semimetals, semiconductors and insulators, adding topological insulators and 3D Dirac semi-metals. These additions represent distinct electronic materials types. Topological insulators have been described in detail in many reviews (see, e.g. \cite{hasan2010colloquium,cava2013crystal,doi:10.7566/JPSJ.82.102001}), In 3D Dirac semi-metals, the conduction and valence bands touch each other at single, discrete points in k-space, around which the dispersion has a characteristically linear k-dependence. In this way, these materials can be seen as a truly 3D analog of graphene. This unusual band dispersion leads the electrons around the Fermi energy to behave like highly relativistic particles, which contrasts with the usual non-relativistic Schrodinger electrons in most metals and semi-metals. While rules and design principles for topological insulators have been posited previously \cite{muchler2012topological}, currently lacking is a bridge to connect physics and design principles for 3D Dirac semi-metals. The purpose of this paper is twofold. First we discuss currently known Dirac semimetals, including briefly those solid solution materials that are expected to achieve this state in a narrow compositional crossover regime between topological and trivial semiconductors (e.g. TlBiSe$_{2-x}$S$_{x}$), and, primarily, the currently known materials that are intrinsic Dirac semimetals (e.g. Cd$_3$As$_2$ and Na$_3$Bi) at their thermodynamic compositions. These latter materials are the focus of the first part of the paper. We then describe three distinct ways of achieving the 3D Dirac semi-metal state from a combined physics and chemistry perspective, along with predictions of new 3D Dirac semimetals based on the presented rationale. The paper is organized such that each section and discussion is independent, as it turns out that the ideas behind each of the three presented ways of realizing a 3D Dirac semi-metal are distinct.

\section{Current Materials}

The first experimentally realized Dirac semi-metal, graphene, has been shown to have a 2D electron gas of Dirac fermions at the Fermi level. This is due to the unique structure and \textit{sp}$^2$ bonding network of the carbon atoms on the honeycomb lattice of graphene, which leads to one unpaired p$_z$ electron per site. Because of this, graphene has six 2D Dirac cones, each located at a K point of the hexagonal Brillouin zone (Figure 1a) \cite{novoselov2005two}. This leads to many interesting transport properties, including very high electron mobilities and chiral quantum Hall effects \cite{geim2007rise}.

After the explosion of research in graphene, interest was piqued in realizing this state in a 3D system. It was known that at the transition between a trivial insulator and topological insulator, a 3D Dirac semi-metal should emerge as an intermediate state\cite{yang2014classification}. An example of this, TlBiSe$_{2-x}$S$_x$, which is between the topological insulator TlBiSe$_2$ and the trivial insulator TlBiS$_2$, has a Dirac point at the Brillouin zone center, or $\Gamma$ point (Figure 1(b))\cite{novak2014large,singh2012topological}. However this is only true for a precise value of x in the solid solution (around x=0.5). It has also been shown experimentally for Hg$_{1-x}$Cd$_x$Te\cite{orlita2014observation}. The 3D Dirac semimetal state is also expected, although not yet experimentally verified, for specific x values in the systems Bi$_{2-x}$In$_{x}$Se$_3$, Pb$_{1-x}$Sn$_x$Se, and Bi$_{1-x}$Sb$_{x}$. In each of these situations, one end member is a trivial insulator, and the other is a topological insulator.

Given that solid solutions are in general not locally homogenous, realizing a true and stable 3D Dirac material in these systems is a challenge. In Bi$_{2-x}$In$_{x}$Se$_3$, a small amount of In is expected to cause the transition from topological insulator to 3D Dirac semi-metal, although the In is also expected to segregate in the material, leading to possible complications \cite{liu2013topological}. In Hg$_{1-x}$Cd$_x$Te \cite{orlita2014observation} it was possible due to the achievement of an extremely homogenous mixing of Hg and Cd. Whether or not it is generally possible in other systems remains to be experimentally verified. However, in spite of the materials challenges, it has been shown that at least near the critical Dirac semimetal composition, anomalous transport properties such as large linear magnetoresistance \cite{novak2014large} and a massive Dirac spectrum (that is, a Dirac cone with a small gap) \cite{liang2013evidence} can be observed, in TlBiSe$_{2-x}$S$_x$ and Pb$_{.77}$Sn$_{.23}$Se respectively. In fact, Pb$_{.77}$Sn$_{.23}$Se undergoes a phase transition between topological and trivial semiconducting states upon cooling, suggesting that other compounds near the critical composition can also change their topology upon cooling or heating due to small changes in the lattice parameters. As such, alternative types of 3D Dirac semi-metals that are not so sensitive to temperature, pressure and composition homogeneity are desired.

Two more recently studied materials, Cd$_3$As$_2$ and Na$_3$Bi. have intrinsic 3D Dirac points along a high symmetry line in the Brillouin zone, protected by an element of crystalline symmetry (C$_4$ and C$_3$, respectively) shown in Figure 1(c-d). These crossing points are robust and are not dependent on temperature or a specific compositional parameter that may vary slightly from point to point in a single crystal \cite{wang2013three,wang2012dirac,PhysRevLett.113.027603,neupane2014observation,liu2014discovery,liu2014stable} . As such, both of these materials represent stable, robust 3D Dirac semi-metals.

In Cd$_3$As$_2$, a band inversion between the empty Cd \textit{s} orbitals and the filled As \textit{p} orbitals provides the nontrivial topology. However unlike topological insulators, the C$_4$ rotation axis prevents the formation of a full band gap in the electronic structure; a Dirac-like band crossing in the electronic structure is protected along the line $\Gamma$-Z in the Brillouin zone. As such, two Dirac points at $\pm$ k$_z$ are observed in this material \cite{wang2013three,wang2012dirac,PhysRevLett.113.027603,neupane2014observation,liu2014discovery}.(Figure 1(c)). This Dirac semimetal has already been shown to have anomalous transport properties, including extremely high electron mobility (larger than that of graphene) and very large magnetoresistance \cite{jeon2014landau,feng2014large, liang2014ultrahigh}. One disadvantage of Cd$_3$As$_2$ is that the band inversion energy, or the energy range over which the electron dispersion is purely "Dirac-like," is very small. Furthermore, crystals of Cd$_3$As$_2$ have a strong tendency to cleave along the pseudo-close-packed (112) crystal plane, which does not preserve the C$_4$ symmetry. This is important in that some proposed experiments require the application of a magnetic field along the C$_4$ rotation axis (the crystallographic [001] direction) \cite{jeon2014landau}. In addition, care is required in the handling of Cd and As during materials synthesis. These factors complicate research into this material, but it can be expected that future studies will continue to unearth its unusual properties.

For Na$_3$Bi, a band inversion of unoccupied Na \textit{s} orbitals and occupied Bi \textit{p} orbitals leads to its non-trivial topology. In this case, the three- and six-fold symmetry axes protect the Dirac points along the c direction, leading again to two Dirac points at $\pm$ k$_z$, although this time in a hexagonal lattice (Figure 1(d)). These Dirac cones have been shown to be highly anisotropic, which is different from the case of Cd$_3$As$_2$ \cite{wang2012dirac,liu2014discovery}. Na$_3$Bi also has a much larger band inversion energy than Cd$_3$As$_2$. This leads to more leeway in a real material for the defects to be controlled such that the Fermi level intersects a pure Dirac dispersion regime of the electronic structure. Also advantageously, crystals of Na$_3$Bi can be cleaved to expose the (001) plane, as well as the (100) plane, allowing for experimental versatility in the probing of the Dirac semi-metal state \cite{xu2013observation}. Unfortunately, however, hopefully obviously to the reader, the growth and care of single crystals of a material that is nearly completely elemental sodium poses significant challenges; Na$_3$Bi is incredibly unstable upon exposure to air and moisture.

Given the experimental difficulties with the current materials, theoretical prediction and experimental validation of new materials is of high interest, and has indeed been called for \cite{zhu2014condensed}. Other Dirac materials exist as well; Dirac materials with a Dirac mass (i.e. the Dirac spectrum has an energy gap) have been known for some time, with Bi being a famous example. A recent example in the AMn$_2$Bi$_2$ family also shows anomalous transport properties\cite{park2011anisotropic}. Proposed but not yet observed Dirac materials with a small mass include the cubic antiperovskites such as Ca$_3$SnO---these compounds, given that they have a bulk gap, are intrinsically related to topological insulators \cite{sun2010new,hsieh2014topological2}. Finally, Dirac-like states have been observed in the Fe based superconductors \cite{harrison2009dirac, morinari2010topological, richard2010observation}. However, these Dirac cones exist within the spin density wave phase, and thus are difficult to treat theoretically. It is further unclear whether they have a small Dirac mass or not. Further work on these materials will be of interest.

The remainder of this paper will focus on the general design principles for finding new 3D Dirac semi-metals, as well as specific examples and predictions for new materials. These design principles are divided into three categories; Case I: 3D Dirac semi-metals derived from charge balanced chemical formulas, Case II: 3D Dirac semi-metals derived by analogy to graphene, and Case III: 3D Dirac semi-metals derived from screw axes and glide planes. For simplicity of discussion, only compounds with inversion symmetry are considered here, as compounds that lack inversion symmetry have complications due to spin splitting of the bands. This can in general lead to a Weyl semi-metal state, which will not be discussed here.

\section{Case I: 3D Dirac semimetals from charge balanced semiconductors}

This section concerns 3D Dirac semi-metals (DSMs) derived from materials whose chemical formulas are related to those of charge balanced semiconductors. Two of the predicted and experimentally verified 3D DSMs, Na$_3$Bi (3 Na$^+$, Bi$^{3-}$) and Cd$_3$As$_2$ (3 Cd$^{2+}$, 2 As$^{3-}$) have this characteristic. Indeed, for both compounds, when the pnictide is replaced by the element one a row up on the periodic table (e.g. Na$_3$Sb and Cd$_3$P$_2$), an actual semiconductor is obtained; this is due to the increased electronegativity difference between the formal cation and anion, as well as the reduced band dispersion that comes from weaker cation-anion covalency. This implies that somewhere between the two end member compounds (i.e. in the solid solutions Cd$_3$P$_{2-x}$As$_x$ and Na$_3$Bi$_{1-x}$Sb$_x$) the band gap can be traced and the composition where the gap closes and the system becomes a 3D Dirac semi-metal can be determined. This is similar but distinct from the case of the gap closing that happens in a trivial insulator to topological insulator transition, as in the system TlBiS$_{2-x}$Se$_x$, where the 3D DSM occurs at a specific value of x between topologically trivial and topological insulator states; for Cd$_3$As$_2$ and Na$_3$Bi the end member material is a 3D DSM.

Going backwards in the narrative somewhat, overall, when a direct band gap closes and the conduction and valence bands overlap, one of three things can happen in a material (Figure 2a): 1. The crossing points are ``gapped out'' due to hybridization (an ``avoided crossing'') and the system becomes a topological insulator with a band inversion. 2. The same as in 1, but the system becomes a normal insulator with a band inversion and 3. The crossing points gap out at all points except certain special points along certain lines of crystal symmetry, leading to a 3D Dirac semi-metal state. \cite{yang2014classification}

What controls which of these scenarios occurs in a real material? The answer lies in the underlying point group symmetry of the crystal structure. The essential idea is this: given the symmetry of the crystal structure and the nature of the orbitals making up the electronic states that cross when the band gap closes, the electronic states must be orthogonal to each other in order to not interact with each other and gap out. To use the familiar point group of the water molecule, C$_{2v}$, as an example, there are four irreducible representations. Electronic states with the same irreducible representation will gap out; ones with different irreducible representations will not. The presence of the spin orbit coupling (SOC) interaction complicates the picture in the materials of interest here. SOC essentially allows for orbitals that would normally be orthogonal to each other (for example, the \textit{s} orbital and \textit{p}$_z$ orbital on the oxygen in H$_2$O in C$_{2v}$) to interact with each other.

The rules of which orbitals truly are orthogonal to each other in the presence of SOC are contained within the "double group", as opposed to the point group \cite{PhysRev.96.280,PhysRev.100.580}. The double group is essentially the point group that also takes into account that the states in question are spin 1/2 particles. For example, the C$_{2v}$ double group only has one irreducible representation, meaning that in this system all states have the same symmetry \cite{koster1963properties}. This is a drastic difference compared to C$_{2v}$ without SOC. This is why SOC has the tendency to gap out band crossings. The strength of this interaction is partially controlled by the atomic number of the atoms involved, as it scales roughly with Z$^4$. 

Therefore, in order to realize the 3D DSM state, at least along some line of crystal symmetry, the states that cross must have different symmetries in the double group. This can be realized in general with C$_3$, C$_4$ and C$_6$ rotation symmetries \cite{koster1963properties}. As the bands can cross at an arbitrary point on all of the lines of crystal symmetry, the full double group isn't used, as that corresponds to the symmetry at the $\Gamma$ point. Instead, the double group of a line of symmetry is used. In this way, not all crystal structures and space groups are equal. For example, in the tetragonal space group I4/mcm, the line along the c axis has C$_{2v}$ symmetry, and therefore cannot have a Dirac point, while in the tetragonal space group P4/mmm, the same line has C$_{4v}$ symmetry, and thus a Dirac point is allowed. The lines along the rotation axes for Cd$_3$As$_2$ (I4$_1$/acd) and Na$_3$Bi (P6$_3$/mmc) have C$_{4v}$ and C$_{6v}$ symmetries, respectively, which allows them to display a 3D Dirac point. One important result of these considerations from the materials perspective is that this type of DSM is not possible in orthorhombic, monoclinic, or triclinic space groups due to the lack of appropriate double group symmetries. This rules out many potential materials candidates in the search for new DSMs.

It is important that the presence of a double group that allows for different irreducible representations does not guarantee that the valence band and conduction bands in the vicinity of the Fermi energy specifically will indeed have different irreducible representations. For example, in Bi$_2$Se$_3$, which along the k$_z$ axis has the symmetry C$_{3v}$, different symmetries are allowed for the bands, and it could have a Dirac point. However, the valence and conduction band orbitals near E$_F$ in fact happen to have the same C$_3$ symmetry, and thus a full band gap opens, forming a topological insulator. The overall formula of these materials, however, can lead to useful predictions. In general, one wants to look for materials that have a gap closing due to the presence of heavy elements and hexagonal, rhombohedral, tetragonal, or cubic symmetry.

A simple family of materials that exhibits these characteristics, including the different symmetry along the c axis necessary to have the Dirac point, is the 111 family of hexagonal ZrBeSi-type compounds. These materials crystallize in the same space group as Na$_3$Bi (P6$_3$/mmc)
\cite{xie2014pressure} and have a very simple crystal structure, with layers of BN-type nets separated by large cations. An archetypal example, BaAgBi (which is charge balanced at Ba$^{2+}$Ag$^{+}$Bi$^{3-}$) is shown in Figure 2(a-b) \cite{kang2002intermetallic}. In this compound and many related ones, the orbitals making up the valence and conduction bands have different symmetries under C$_3$ rotation in the double group. This allows for gapping out by SOC of all states, with the exception of one point along $\Gamma$-A, leading to a 3D DSM state, as shown in Figure 2(d). Many materials in this family have this 3D Dirac point. There is often the complication, however, of the presence of other, non-Dirac bands; this is the case for SrAgBi, shown in Figure 2(e)).

The LiGaGe-structure family, related to the above materials but with a puckering of the honeycomb net, also contains Dirac cones in the electronic structure. Although this system does not have inversion symmetry (the space group is P6$_3$mc), the states are not spin split along the relevant $\Gamma$-A line. Thus these materials can also be 3D DSMs, an example of which, YbAuSb (Figure 2(c)), is shown in Figure 2(f).This family is chemically convenient as it is can host a large number of different elements, including magnetic f$^7$ Eu$^{2+}$. This sort of band structure tuning through changing Z in a large chemical family also occurs in the half-Heusler compounds, which become topological insulators\cite{chadov2010tunable}. There are also further related hexagonal families, such as the four-layer LaAuSb family, that have these Dirac cones along $\Gamma$-A.

As such, a more detailed classification is possible in the 111 ZrBeSi type materials family. We show in Figure 3a how the electronic characteristics of the family change as a function of the total Z divided by the Pauling electronegativity difference (we use the electronegativity difference between the large cation and the average of the anionic honeycomb sublattice). The figure shows whether or not the system has a Dirac point, and how much other bands are interfering, quantified by the density of states (DOS) at the Fermi level (E$_F$) (a perfect 3D DSM should have essentially zero DOS at E$_F$). While not completely systematic, it can be seen that once this Z/E metric reaches a certain value, the compounds have a Dirac cone in their electronic structure (Figure 3(a)).
 
We also show this evolution more simply as a function of Z (Figure 3(b)). Here, Z correlates well with the DOS. This is because a larger Z leads to a larger SOC which in turn results in a better gapping of the system. (This can be seen by comparing BaAgBi to SrAgBi.) The calculated DOS for all the compounds is likely to be higher than in the real materials, as the PBE functional tends to underestimate band gaps, but the trend is very clear. In this structural family the compound BaAgBi appears to be the most promising potentially new 3D Dirac semimetal. Its study experimentally would be of significant interest.

Because there exists a C$_4$ protected tetragonal 3D DSM (Cd$_3$As$_2$) as well as C$_3$ and C$_6$ protected 3D DSMs (Na$_3$Bi and proposed in the ZrBeSi family), it is natural to ask whether this can be realized in a cubic system that can has C$_3$ symmetries (as the C$_4$ symmetry is not required in cubic materials). Indeed, our calculations on the pyrite MX$_2$ family show that the heavy compound PtBi$_2$(Figure 4(a)), space group Pa$\bar{3}$(which is isoelectronic and isostructural to the semiconductor PtSb$_2$)\cite{brese1994bonding,} has a 3D Dirac point along the line $\Gamma$-R, which falls along the threefold rotation axis with symmetry C$_{3v}$ (Figure 4(c)). While other bands cross the Fermi level, there is still a continuous gap, and this serves as proof of concept that this type of 3D DSM is possible in cubic systems. 
 
As a final note, it is theoretically possible that a stoichiometric compound naturally realizes the topological critical point criterion for the 3D DSM, without necessarily being protected by a rotation symmetry. While in TlBiSe$_{2-x}$S$_x$ this occurs at a critical value of x, our calculation of the electronic structure of the compound SrSn$_2$As$_2$(Figure 4(b)), which is isostructural to Bi$_2$Te$_2$Se \cite{villars2008srsn2as2}, shows that the compound is naturally very near the critical point. This suggests that even without tuning it may realize the 3D DSM state (Figure 4(d)).

Thus here we described the general principles behind predicting 3D DSMs in charge balanced semiconductors based on symmetry considerations, and predict specifically that materials in the ZrBeSi materials family, specifically BaAgBi; the LiGaGe materials family, specifically YbAuSb; the pyrite family, specifically PtBi$_2$; and the Bi$_2$Te$_2$Se family, specifically SrSn$_2$As$_2$; are of interest as potentially new 3D Dirac semimetals.      

\section{Case II: 3D Dirac semimetals from orbital degeneracies}

Besides the band overlap in compounds with charge balanced semiconducting formulas, there are other ways of achieving the Dirac semimetal state. In fact the most famous 2D Dirac semi-metal is graphene, which is based on a different, but related concept, in which certain orbitals are forced to be degenerate at certain points in the Brillouin zone due to crystal symmetry. In graphene, this is rather simple, and has been explored previously. Essentially, as there are two C atoms in a honeycomb array per unit cell, there are two p$_z$ orbitals per unit cell. As such, two basis sets, a bonding one and an anti-bonding one, construct the band structure. Due to the crystal symmetry, these two basis sets become equivalent at the K point, which forces them to be degenerate. This leads to a 2D DSM state. This degeneracy is susceptible to gapping out due to SOC, but in graphene the SOC effect is not strong enough to do this due to the low Z. The stacking of layers on top of each other also ruins this effect, meaning that the 3D material graphite is no longer a DSM.

So how can the same concepts that make graphene a DSM create a 3D material that is also a DSM? As the stacking of the honeycomb layers is what gaps out the Dirac point in graphite, we suggest that moving the layers further apart in a 3D material so that their interaction is very weak can either reduce or eliminate this gap. This can be done by inserting a layer of positive cations between the honeycomb layers while maintaining the same electron count as graphene. This occurs in MgB$_2$,for example, in which the B$_2$$^{-2}$ layer has the same structure as graphene, and the layers are separated by layers of Mg$^{2+}$ ions. However, while MgB$_2$ is metallic and in fact superconducting \cite{buzea2001review}, the interlayer coupling is still too strong as there is too much covalent bonding between Mg and B. In order to regain the Dirac point at K, the interlayer coupling must be reduced even more, by going to larger cations.

Following this logic, our calculations show that the compound BaGa$_2$ (Figure 5(a)) \cite{iandelli1955structure}, which contains layers of very large and very ionic Ba atoms, retains a Dirac cone centered at K, and very little interlaying coupling. This leads to a very small gap at H, which is the wave vector above K in 3D. The cone comes from the Ga \textit{p}$_z$ orbitals, which have a very small SOC interaction, and therefore create a negligible gap (Figure 5(b)). As such, this can be described as a quasi 2D Dirac cone. In BaGa$_2$, unfortunately, there are other bands present at the Fermi level as well, so it is not a perfect candidate material. The concept, however, remains valid. This concept also applies to the material Bi$_{14}$Rh$_9$I$_3$ in which a graphene-like layer of Bi-Rh cubes is separated by very large spaces in the z direction \cite{rasche2013crystal}. This leads to Dirac points at various K points (as the compound is not hexagonal) that are, however, gapped out by SOC due to the high atomic weight of the atoms in question. 

Another way to avoid the problem of 3D stacking is by having the Dirac point come from orbitals that do not interact with each other along the c axis. For example, a Dirac point in 2D at K can be also derived from \textit{p}$_x$ and \textit{p}$_y$ orbitals, which is the case for the hypothetical layered compound BiI where the p$_z$ orbitals are bonded to a ligand that locks them out \cite{song2014quantum}. (In compounds of this type, the SOC interaction is large and gaps out the system.) In 3D, this type of electronic system is possible to achieve in compounds with hexagonal layers of atoms that are in an overall linear coordination. One example of this is the material PbTaSe$_2$. In this case, the 2D Pb layer is in linear coordination with the Se atoms, leading to a hexagonal lattice of \textit{p}$_x$ and \textit{p}$_y$ orbitals, which creates a Dirac point in the Pb-derived electronic structure at K \cite{PhysRevB.89.020505}. This Dirac point is still there in the full 3D electronic structure. However, as it is from Pb p orbitals, the SOC interaction is very large and creates a very large gap. Furthermore, The TaSe$_2$ sublattice contributes bands, and the resulting compound is superconducting. However, lighter elements in this sort of configuration, such as a hypothetical SiTaSe$_2$, could create another quasi-2D DSM.

Another example of hexagonally ordered atoms in linear coordination can be found in a couple of platinum-based materials in which there are linear chains of platinum anions that are ordered hexagonally in 2D. These include BaPt and Li$_2$Pt, in which Pt$^{2-}$ anions are in linear coordination with each other \cite{karpov2004covalently,bronger1975synthesis}. The crystal structure of BaPt is shown in Figure 5(c), highlighting the Pt chains.  This leads to a Dirac point at K for the case of no SOC for both materials. However, again, the SOC present in the real materials gaps them out. This is shown in Figure 5(d). However, the concept that linear chains of atoms that are ordered hexagonally in 2D can lead to a Dirac point at K, as in graphene, holds. In the case of linearly coordinated Pt$^{2-}$ ions, it is the in-plane \textit{d}$_{x^2-y^2}$ and d$_{xy}$ orbitals that combine to form the Dirac cone, similar to how in PbTaSe$_2$ the it is the in-plane\textit{p}$_x$ and \textit{p}$_y$ orbitals, that do so.

Thus we argue that 3D DSMs are possible for this kind of materials system, where hexagonally-based sheets of atoms form the basis of the electronic structure, in analogy to graphene, but where the bonding is such that the interactions in the third dimension are weak. We have not yet been able to identify a real material that is an excellent candidate example, but show that BaGa$_2$, Bi$_{14}$Rh$_9$I$_3$, PbTaSe$_2$, Li$_2$Pt and BaPt are real materials that may display interesting electronic properties as a result of displaying these types of electronic states. 

\section{case III: 3D Dirac semimetals from glide planes and screw axes}

The effects of "non-symmorphic" symmetry, or space groups containing glide planes and/or screw axes, on the electronic band structure of solids has been known for a long time \cite{zak2001topologically}. The essence is that the presence of these symmetry elements makes it such that the normally two-fold degenerate bands must touch at four fold degenerate points at certain special points at the surface of the Brillouin zone, and that some of these degeneracies are not susceptible to SOC\cite{PhysRevB.59.5998}. One can naturally see how this can be used to create the 3D DSM state. However, due to the pairing of the bands, these sorts of 3D DSMs cannot be created from charge balanced formulas, or as natural extensions of semiconducting formulas, as was the case for Na$_3$Bi and Cd$_3$As$_2$. As has been discussed previously, in non-symmorphic structures an odd number of electrons per formula unit cannot result in a normal insulator \cite{parameswaran2013topological}. This can be rationalized chemically. In order for a crystal structure to be non-symmorphic, there must be at least a doubling of the formula unit per unit cell. For example, the simplest non-symmorphic semiconductor is elemental Si (space group Fd-3m), which contains two Si atoms per unit cell that are related to each other by a glide plane. Because of this, the electronic bands come in pairs, and must touch at certain points on the Brillouin zone surface. Each Si has four electrons, so there are 8 electrons in total, or four filled bands per unit cell. This means that two pairs of bands are filled, which in turn means that a true band gap can be achieved. In order to have E$_F$ at a "Dirac point", there would have to be a half-filled pair of bands. This necessitates an odd number of electrons per formula unit. A general schematic of this scenario is shown in Figure 6(a).

This is exactly the logic that led to the theoretically predicted 3D DSM BiO$_2$ in the SiO$_2$ structure, a hypothetical (and never to be synthesized) compound \cite{young2012dirac}. The cubic SiO$_2$ structure contains a diamond lattice of Si$^{4+}$,analogous to that in elemental Si, with SiO$_2$ itself having an even number of electrons per site, therefore resulting in a normal insulator. Hypothetical BiO$_2$ would have one more electron per formula unit with a diamond lattice of Bi$^{4+}$ (or one electron per site) and therefore two more electrons per unit cell, such that one more band is filled, or one half of a pair of bands is filled. Thus, the Fermi energy would coincide with one of the "sticking points" in the electronic structure. In order for other bands to not interfere, one would like the bands that are half filled to be far in energy from any other bands. This occurs in hypothetical BiO$_2$ wherein the half-filled Bi \textit{6s} shell is far removed in energy from any other orbitals. However, a half filled 6s orbital is unstable \cite{schoop2013lone} and in any case Bi$^{4+}$ is too large to ever have tetrahedral coordination with oxygen. \cite{schoop2013lone}. Similarly, the (impossible to ever synthesize) hypothetical spinel structure compound BiAl$_2$O$_4$ has also been proposed to display this type of electronic system \cite{steinberg2014bulk}. 

In order to achieve this sort of 3D DSM in a real material, a stable compound with an odd number of electrons per site, in an orbital manifold that is far removed from other orbitals in energy, must be synthesized. Main group elements do not tend to form oxidation states with odd numbers of electrons, and states such as Bi$^{4+}$ are highly unstable. Transition metals can form such electronic states, however, although they tend to have d orbitals that lay very close to each other in energy. One chemically stable materials family based on a transition metal with an odd number of electrons per site that is a proposed 3D DSM derived from non-symmorphic symmetry are the Ir$^4+$ pyrochlores. The Ir$^4+$ present has a half filled \textit{J}$_{1/2}$ orbital, which is far enough from the other d orbitals in energy due to the large SOC of Ir \cite{wan2011topological,PhysRevB.84.075129}. In this case, and for transition metal-based systems in general, localized, magnetic electrons are often found, which poses challenges for understanding whether a delocalized, conducting 3D DSM state can actually be realized. 

There are two possible solutions to this problem of achieving an effective one electron per site lattice. One is that instead of having one electron per atomic orbital, which tends to be either chemically unstable or highly localized, one can use one electron per molecular orbital, as often occurring in cluster compounds. One example of a material where this may work is the compound TlMo$_3$Te$_3$, or in fact the whole AMo$_3$X$_3$ family, where A=(Na, K, Rb, In, Tl), X=(Se,Te) (Figures 6(b-c) \cite{chevrel1985ternary}. In these compounds, the crystal structure is that of condensed chains of face sharing Mo$_6$ octahedra, surrounded by X, with A filling the channels in between the chains. The face sharing octahedra can be considered as Mo$_3$ triangles related to each other by a screw axis; as such the crystal structure is non-symmorphic. Each Mo$_3$ triangle has one extra electron donated to it by the A$^+$ ion, which leads to the band structure shown in Figure 5(d). One can clearly see the sticking points in the electronic structure, indicative of half filling of a molecular orbital. Unlike the iridate pyrochlores, the states are delocalized due to the closesness of the Mo$_3$ triangles to each other. Critically from an experimental perspective, these materials are chemically stable. Due to the chain nature of the crystal structure of the compounds, the Dirac state is expected to be quasi-1D rather than fully 3D (Figure 6(d). Other cluster compounds may also be expected to show similar effects. This idea is roughly analogous to the chemical concept that atomic radicals are very unstable, but that radicals can be stabilized by delocalizing the electron over a larger molecular orbital. As such, as shown in this family, cluster compounds may be one promising way of achieving the non-symmorphic 3D DSM state. 

A family of compounds with a very similar electronic structure to this is the family of group 4 trihalides, for example HfI$_3$ \cite{struss1969lower}. These have linear chains of d$^1$ transition metal ions, where the ions are close enough to be considered delocalized. The electronic structure that we calculate for this material is shown in Figure 6e. The calculated Dirac cone in HfI$_3$ is even more quasi-1D than that calculated for TlMo$_3$Te$_3$. Indeed, the isostructural and chemically analogous material ZrI$_3$ has been shown to undergo a Peierls-like distortion in which Zr-Zr dimers are formed, gapping out the electronic states \cite{lachgar1990revision}. Peierls distortions are another possible instability to be wary of for electronic systems with 1 electron per site. This is also the case for the \textit{d}$^1$ group 5 compounds such as NbSe$_2$ and TaSe$_2$ which also have one electron per site; these compounds are famous for their various electronic instabilities including Charge Density Wave (CDW) formation and superconductivity. 

Cluster compounds may be a way to minimize these instabilities as well, although the compound TlMo$_3$Se$_3$ is reported to be superconducting\cite{lepetit1984superconductivity}. In fact, these quasi 1D Dirac compounds may prove a fertile ground for interesting physics, as our calculations indicate that CDW (or Peierls) and superconducting instabilities occur within the 3D Dirac state band manifold, implying that these materials represent "interacting" 3D Dirac electrons. These systems are not perfect 3D DSMs, however, due to having more than one sticking point in the Brillouin zone---thus cluster compounds or early transition metal compounds with different symmetry may prove more ideal.

Another solution to the problem lies in intermetallic systems, in which a simple oxidation state is difficult to assign. In such a case, the Fermi level can certainly be tuned to a non-symmorphic "Dirac point". In intermetallic systems, however, there tend to be multiple bands crossing E$_F$. One example is the paramagnetic metal Cr$_2$B \cite{schoop2014paramagnetic}. Cr$_2$B is in the non-symmorphic space group Fddd, and contains interlocking honeycomb-like nets of Cr atoms related to each other by glide planes (Figure 7(a)). Cr$_2$B has no obvious apparent formal oxidation state for Cr, and as a true intermetallic, has many bands crossing the Fermi energy, all of which contribute to the charge transport. However, in the band structure, the non-symmoprhic "sticking points" formed from linear bands can be easily identified (Figure 7(b)). Given that Cr$_2$B can have a variable B content, we show in the figure the calculated band structure for both stoichiometric Cr$_2$B and Boron deficient Cr$_2$B$_{0.9}$ (Figure 7(c)). At Cr$_2$B$_{0.9}$, the Dirac point comes to E$_F$. While there are many other bands as well, Hall effect measurements on polycrystalline Cr$_2$B samples uncovered n-type carriers with high mobility whose effects can be distinguished from those of the other carriers (Figure 7(d)). This indicates that it is possible that even if other bands cross E$_F$, the Dirac electrons can still be visible in transport experiments due to their high mobility.

Finally, among this third class of potential 3D Dirac Semimetals, we identify the specific materials HfI$_3$ and TlMo$_3$Te$_3$ as worthy of future study, and argue that even for classic intermetallic phases such as Cr$_2$B, the effects of non-symmorphic symmetry can result in the presence of Dirac electrons whose contribution to the transport properties may be observable. 

\section{Methods}
Electronic structure calculations were performed in the framework of density functional theory using the Wien2k code\cite{Blaha1990399} with a full-potential linearized augmented plane-wave and local orbitals basis together with the Perdew-Burke-Ernzerhof parameterization of the generalized gradient approximation, or the modified Becke Johnson exchange functional, as indicated\cite{becke:221101,perdew1996generalized}.  The plane wave cutoff parameter R$_{MT}$K$_{max}$ was set to 7 and the Brilloun zone (BZ) was sampled by 2000 k-points. The mBJ functional was used due to its improved accuracy for band gaps, and also for improved band inversion strengths for band-inverted systems such as 3D Dirac semi-metals .

Cr2B$_{.95}$ samples were synthesized as published elsewhere\cite{schoop2014paramagnetic}. The Hall effect and resistivity of Cr$_2$B were measured on poly-crystalline platelets of approximate dimensions $2 \times 6\times 0.08\,$mm in a standard six-wire geometry, using our in-house cryostat and magnet. The current was applied along the longest edge, while the magnetic field was perpendicular to the sample plane. We fitted the Hall conductivity $\sigma_{xy}$ to a two-band model which combines the full Drude expression $en\mu^2B/(1+(\mu B)^2)$ for the fast band with a linear term $c B$, where the parameter $c$ is expected to include several (linearized) contributions from the bands of low carrier mobility. Here $e$, $n$, and $\mu$ are the fundamental charge of the electron, the carrier density of the fast band and its mobility, respectively. The resulting carrier mobility of the fast band is approximately independent of temperature below $40\,$K.

\section{Conclusions}

This paper has firstly presented a discussion of currently known 3D Dirac materials, including those that show such behavior at a specific composition in a solid solution system, such as TlBiSe$_{2-x}$S$_x$, and those that are native 3D Dirac semimetals such as Cd$_3$As$_2$ and Na$_3$Bi. Consideration of those materials, graphene, and other factors allowed us to propose general design principles that will be of use for finding new 3D Dirac semimetals, and present predictions for specific new materials. Each design principle leads to unique predictions that warrant future research. For case I, Dirac materials based on charge balanced formulas, it was posited that the ZrBeSi family of materials (such as SrAgBi and BaAgBi), the LiGaGe family of materials (such as YbAuSb) and other materials such as PtBi$_2$ and SrSn$_2$As$_2$ as compounds should display 3D Dirac cones. For case II, Dirac materials in analogy to graphene, the prediction of a 3D Dirac cone in BaAgBi was made. For case III, Dirac materials based on glide planes and screw axes, analysis led to the identification of AMo$_3$X$_3$, HfI$_3$ and Cr$_2$B as compounds with quasi 1D Dirac cones in the bulk, and a Dirac cone buried with other bands in an intermetallic, respectively. Given the relative novelty of the field of 3D Dirac semi-metals, these predictions may lead to not only 3D Dirac materials more suitable for experimental study, but also ones with magnetic, superconducting, and CDW instabilities, which should expand the field of 3D Dirac semi-metals. It is also expected that the design principles described here will lead to identification of additional candidate materials, opening up a new area of research in this emerging field.

\bigskip 
\begin{acknowledgments}

The authors' research on Dirac semimetals is funded by the ARO MURI on topological insulators, grant W911NF-12-1-0961, and the NSF funded MRSEC at Princeton University, grant DMR-0819860.

\end{acknowledgments}
\bibliography{Dirac_bib_2}

\begin{thebibliography}{61}%
\makeatletter
\providecommand \@ifxundefined [1]{%
 \@ifx{#1\undefined}
}%
\providecommand \@ifnum [1]{%
 \ifnum #1\expandafter \@firstoftwo
 \else \expandafter \@secondoftwo
 \fi
}%
\providecommand \@ifx [1]{%
 \ifx #1\expandafter \@firstoftwo
 \else \expandafter \@secondoftwo
 \fi
}%
\providecommand \natexlab [1]{#1}%
\providecommand \enquote  [1]{``#1''}%
\providecommand \bibnamefont  [1]{#1}%
\providecommand \bibfnamefont [1]{#1}%
\providecommand \citenamefont [1]{#1}%
\providecommand \href@noop [0]{\@secondoftwo}%
\providecommand \href [0]{\begingroup \@sanitize@url \@href}%
\providecommand \@href[1]{\@@startlink{#1}\@@href}%
\providecommand \@@href[1]{\endgroup#1\@@endlink}%
\providecommand \@sanitize@url [0]{\catcode `\\12\catcode `\$12\catcode
  `\&12\catcode `\#12\catcode `\^12\catcode `\_12\catcode `\%12\relax}%
\providecommand \@@startlink[1]{}%
\providecommand \@@endlink[0]{}%
\providecommand \url  [0]{\begingroup\@sanitize@url \@url }%
\providecommand \@url [1]{\endgroup\@href {#1}{\urlprefix }}%
\providecommand \urlprefix  [0]{URL }%
\providecommand \Eprint [0]{\href }%
\@ifxundefined \urlstyle {%
  \providecommand \doi  [0]{\begingroup \@sanitize@url \@doi}%
  \providecommand \@doi [1]{\endgroup \@@startlink {\doibase
  #1}doi:\discretionary {}{}{}#1\@@endlink }%
}{%
  \providecommand \doi  [0]{doi:\discretionary{}{}{}\begingroup
  \urlstyle{rm}\Url }%
}%
\providecommand \doibase [0]{http://dx.doi.org/}%
\providecommand \Doi [0]{\begingroup \@sanitize@url \@Doi }%
\providecommand \@Doi  [1]{\endgroup\@@startlink{\doibase#1}\@@Doi}%
\providecommand \@@Doi [1]{#1\@@endlink}%
\providecommand \selectlanguage [0]{\@gobble}%
\providecommand \bibinfo  [0]{\@secondoftwo}%
\providecommand \bibfield  [0]{\@secondoftwo}%
\providecommand \translation [1]{[#1]}%
\providecommand \BibitemOpen [0]{}%
\providecommand \bibitemStop [0]{}%
\providecommand \bibitemNoStop [0]{.\EOS\space}%
\providecommand \EOS [0]{\spacefactor3000\relax}%
\providecommand \BibitemShut  [1]{\csname bibitem#1\endcsname}%
\bibitem [{\citenamefont {Hasan}\ and\ \citenamefont
  {Kane}(2010)}]{hasan2010colloquium}%
  \BibitemOpen
  \bibfield  {author} {\bibinfo {author} {\bibfnamefont {M.~Z.}\ \bibnamefont
  {Hasan}}\ and\ \bibinfo {author} {\bibfnamefont {C.~L.}\ \bibnamefont
  {Kane}},\ }\href@noop {} {\bibfield  {journal} {\bibinfo  {journal} {Reviews
  of Modern Physics},\ }\textbf {\bibinfo {volume} {82}},\ \bibinfo {pages}
  {3045} (\bibinfo {year} {2010})}\BibitemShut {NoStop}%
\bibitem [{\citenamefont {Cava}\ \emph {et~al.}(2013)\citenamefont {Cava},
  \citenamefont {Ji}, \citenamefont {Fuccillo}, \citenamefont {Gibson},\ and\
  \citenamefont {Hor}}]{cava2013crystal}%
  \BibitemOpen
  \bibfield  {author} {\bibinfo {author} {\bibfnamefont {R.}~\bibnamefont
  {Cava}}, \bibinfo {author} {\bibfnamefont {H.}~\bibnamefont {Ji}}, \bibinfo
  {author} {\bibfnamefont {M.}~\bibnamefont {Fuccillo}}, \bibinfo {author}
  {\bibfnamefont {Q.}~\bibnamefont {Gibson}}, \ and\ \bibinfo {author}
  {\bibfnamefont {Y.}~\bibnamefont {Hor}},\ }\href@noop {} {\bibfield
  {journal} {\bibinfo  {journal} {Journal of Materials Chemistry C},\ }\textbf
  {\bibinfo {volume} {1}},\ \bibinfo {pages} {3176} (\bibinfo {year}
  {2013})}\BibitemShut {NoStop}%
\bibitem [{\citenamefont {Ando}(2013)}]{doi:10.7566/JPSJ.82.102001}%
  \BibitemOpen
  \bibfield  {author} {\bibinfo {author} {\bibfnamefont {Y.}~\bibnamefont
  {Ando}},\ }\Doi {10.7566/JPSJ.82.102001} {\bibfield  {journal} {\bibinfo
  {journal} {Journal of the Physical Society of Japan},\ }\textbf {\bibinfo
  {volume} {82}},\ \bibinfo {pages} {102001} (\bibinfo {year} {2013})},\
  \Eprint {http://arxiv.org/abs/http://dx.doi.org/10.7566/JPSJ.82.102001}
  {http://dx.doi.org/10.7566/JPSJ.82.102001} \BibitemShut {NoStop}%
\bibitem [{\citenamefont {M{\"u}chler}\ \emph {et~al.}(2012)\citenamefont
  {M{\"u}chler}, \citenamefont {Zhang}, \citenamefont {Chadov}, \citenamefont
  {Yan}, \citenamefont {Casper}, \citenamefont {K{\"u}bler}, \citenamefont
  {Zhang},\ and\ \citenamefont {Felser}}]{muchler2012topological}%
  \BibitemOpen
  \bibfield  {author} {\bibinfo {author} {\bibfnamefont {L.}~\bibnamefont
  {M{\"u}chler}}, \bibinfo {author} {\bibfnamefont {H.}~\bibnamefont {Zhang}},
  \bibinfo {author} {\bibfnamefont {S.}~\bibnamefont {Chadov}}, \bibinfo
  {author} {\bibfnamefont {B.}~\bibnamefont {Yan}}, \bibinfo {author}
  {\bibfnamefont {F.}~\bibnamefont {Casper}}, \bibinfo {author} {\bibfnamefont
  {J.}~\bibnamefont {K{\"u}bler}}, \bibinfo {author} {\bibfnamefont {S.-C.}\
  \bibnamefont {Zhang}}, \ and\ \bibinfo {author} {\bibfnamefont
  {C.}~\bibnamefont {Felser}},\ }\href@noop {} {\bibfield  {journal} {\bibinfo
  {journal} {Angewandte Chemie},\ }\textbf {\bibinfo {volume} {124}},\ \bibinfo
  {pages} {7333} (\bibinfo {year} {2012})}\BibitemShut {NoStop}%
\bibitem [{\citenamefont {Novoselov}\ \emph {et~al.}(2005)\citenamefont
  {Novoselov}, \citenamefont {Geim}, \citenamefont {Morozov}, \citenamefont
  {Jiang}, \citenamefont {Grigorieva}, \citenamefont {Dubonos},\ and\
  \citenamefont {Firsov}}]{novoselov2005two}%
  \BibitemOpen
  \bibfield  {author} {\bibinfo {author} {\bibfnamefont {K.}~\bibnamefont
  {Novoselov}}, \bibinfo {author} {\bibfnamefont {A.~K.}\ \bibnamefont {Geim}},
  \bibinfo {author} {\bibfnamefont {S.}~\bibnamefont {Morozov}}, \bibinfo
  {author} {\bibfnamefont {D.}~\bibnamefont {Jiang}}, \bibinfo {author}
  {\bibfnamefont {M.~K.~I.}\ \bibnamefont {Grigorieva}}, \bibinfo {author}
  {\bibfnamefont {S.}~\bibnamefont {Dubonos}}, \ and\ \bibinfo {author}
  {\bibfnamefont {A.}~\bibnamefont {Firsov}},\ }\href@noop {} {\bibfield
  {journal} {\bibinfo  {journal} {nature},\ }\textbf {\bibinfo {volume}
  {438}},\ \bibinfo {pages} {197} (\bibinfo {year} {2005})}\BibitemShut
  {NoStop}%
\bibitem [{\citenamefont {Geim}\ and\ \citenamefont
  {Novoselov}(2007)}]{geim2007rise}%
  \BibitemOpen
  \bibfield  {author} {\bibinfo {author} {\bibfnamefont {A.~K.}\ \bibnamefont
  {Geim}}\ and\ \bibinfo {author} {\bibfnamefont {K.~S.}\ \bibnamefont
  {Novoselov}},\ }\href@noop {} {\bibfield  {journal} {\bibinfo  {journal}
  {Nature materials},\ }\textbf {\bibinfo {volume} {6}},\ \bibinfo {pages}
  {183} (\bibinfo {year} {2007})}\BibitemShut {NoStop}%
\bibitem [{\citenamefont {Yang}\ and\ \citenamefont
  {Nagaosa}(2014)}]{yang2014classification}%
  \BibitemOpen
  \bibfield  {author} {\bibinfo {author} {\bibfnamefont {B.-J.}\ \bibnamefont
  {Yang}}\ and\ \bibinfo {author} {\bibfnamefont {N.}~\bibnamefont {Nagaosa}},\
  }\href@noop {} {\bibfield  {journal} {\bibinfo  {journal} {arXiv preprint
  arXiv:1404.0754}} (\bibinfo {year} {2014})}\BibitemShut {NoStop}%
\bibitem [{\citenamefont {Novak}\ \emph {et~al.}(2014)\citenamefont {Novak},
  \citenamefont {Sasaki}, \citenamefont {Segawa},\ and\ \citenamefont
  {Ando}}]{novak2014large}%
  \BibitemOpen
  \bibfield  {author} {\bibinfo {author} {\bibfnamefont {M.}~\bibnamefont
  {Novak}}, \bibinfo {author} {\bibfnamefont {S.}~\bibnamefont {Sasaki}},
  \bibinfo {author} {\bibfnamefont {K.}~\bibnamefont {Segawa}}, \ and\ \bibinfo
  {author} {\bibfnamefont {Y.}~\bibnamefont {Ando}},\ }\href@noop {} {\bibfield
   {journal} {\bibinfo  {journal} {arXiv preprint arXiv:1408.2183}} (\bibinfo
  {year} {2014})}\BibitemShut {NoStop}%
\bibitem [{\citenamefont {Singh}\ \emph {et~al.}(2012)\citenamefont {Singh},
  \citenamefont {Sharma}, \citenamefont {Lin}, \citenamefont {Hasan},
  \citenamefont {Prasad},\ and\ \citenamefont {Bansil}}]{singh2012topological}%
  \BibitemOpen
  \bibfield  {author} {\bibinfo {author} {\bibfnamefont {B.}~\bibnamefont
  {Singh}}, \bibinfo {author} {\bibfnamefont {A.}~\bibnamefont {Sharma}},
  \bibinfo {author} {\bibfnamefont {H.}~\bibnamefont {Lin}}, \bibinfo {author}
  {\bibfnamefont {M.}~\bibnamefont {Hasan}}, \bibinfo {author} {\bibfnamefont
  {R.}~\bibnamefont {Prasad}}, \ and\ \bibinfo {author} {\bibfnamefont
  {A.}~\bibnamefont {Bansil}},\ }\href@noop {} {\bibfield  {journal} {\bibinfo
  {journal} {Physical Review B},\ }\textbf {\bibinfo {volume} {86}},\ \bibinfo
  {pages} {115208} (\bibinfo {year} {2012})}\BibitemShut {NoStop}%
\bibitem [{\citenamefont {Orlita}\ \emph {et~al.}(2014)\citenamefont {Orlita},
  \citenamefont {Basko}, \citenamefont {Zholudev}, \citenamefont {Teppe},
  \citenamefont {Knap}, \citenamefont {Gavrilenko}, \citenamefont {Mikhailov},
  \citenamefont {Dvoretskii}, \citenamefont {Neugebauer}, \citenamefont
  {Faugeras} \emph {et~al.}}]{orlita2014observation}%
  \BibitemOpen
  \bibfield  {author} {\bibinfo {author} {\bibfnamefont {M.}~\bibnamefont
  {Orlita}}, \bibinfo {author} {\bibfnamefont {D.}~\bibnamefont {Basko}},
  \bibinfo {author} {\bibfnamefont {M.}~\bibnamefont {Zholudev}}, \bibinfo
  {author} {\bibfnamefont {F.}~\bibnamefont {Teppe}}, \bibinfo {author}
  {\bibfnamefont {W.}~\bibnamefont {Knap}}, \bibinfo {author} {\bibfnamefont
  {V.}~\bibnamefont {Gavrilenko}}, \bibinfo {author} {\bibfnamefont
  {N.}~\bibnamefont {Mikhailov}}, \bibinfo {author} {\bibfnamefont
  {S.}~\bibnamefont {Dvoretskii}}, \bibinfo {author} {\bibfnamefont
  {P.}~\bibnamefont {Neugebauer}}, \bibinfo {author} {\bibfnamefont
  {C.}~\bibnamefont {Faugeras}},  \emph {et~al.},\ }\href@noop {} {\bibfield
  {journal} {\bibinfo  {journal} {Nature Physics}} (\bibinfo {year}
  {2014})}\BibitemShut {NoStop}%
\bibitem [{\citenamefont {Liu}\ and\ \citenamefont
  {Vanderbilt}(2013)}]{liu2013topological}%
  \BibitemOpen
  \bibfield  {author} {\bibinfo {author} {\bibfnamefont {J.}~\bibnamefont
  {Liu}}\ and\ \bibinfo {author} {\bibfnamefont {D.}~\bibnamefont
  {Vanderbilt}},\ }\href@noop {} {\bibfield  {journal} {\bibinfo  {journal}
  {Physical Review B},\ }\textbf {\bibinfo {volume} {88}},\ \bibinfo {pages}
  {224202} (\bibinfo {year} {2013})}\BibitemShut {NoStop}%
\bibitem [{\citenamefont {Liang}\ \emph {et~al.}(2013)\citenamefont {Liang},
  \citenamefont {Gibson}, \citenamefont {Xiong}, \citenamefont {Hirschberger},
  \citenamefont {Koduvayur}, \citenamefont {Cava},\ and\ \citenamefont
  {Ong}}]{liang2013evidence}%
  \BibitemOpen
  \bibfield  {author} {\bibinfo {author} {\bibfnamefont {T.}~\bibnamefont
  {Liang}}, \bibinfo {author} {\bibfnamefont {Q.}~\bibnamefont {Gibson}},
  \bibinfo {author} {\bibfnamefont {J.}~\bibnamefont {Xiong}}, \bibinfo
  {author} {\bibfnamefont {M.}~\bibnamefont {Hirschberger}}, \bibinfo {author}
  {\bibfnamefont {S.~P.}\ \bibnamefont {Koduvayur}}, \bibinfo {author}
  {\bibfnamefont {R.}~\bibnamefont {Cava}}, \ and\ \bibinfo {author}
  {\bibfnamefont {N.}~\bibnamefont {Ong}},\ }\href@noop {} {\bibfield
  {journal} {\bibinfo  {journal} {Nature communications},\ }\textbf {\bibinfo
  {volume} {4}} (\bibinfo {year} {2013})}\BibitemShut {NoStop}%
\bibitem [{\citenamefont {Wang}\ \emph {et~al.}(2013)\citenamefont {Wang},
  \citenamefont {Weng}, \citenamefont {Wu}, \citenamefont {Dai},\ and\
  \citenamefont {Fang}}]{wang2013three}%
  \BibitemOpen
  \bibfield  {author} {\bibinfo {author} {\bibfnamefont {Z.}~\bibnamefont
  {Wang}}, \bibinfo {author} {\bibfnamefont {H.}~\bibnamefont {Weng}}, \bibinfo
  {author} {\bibfnamefont {Q.}~\bibnamefont {Wu}}, \bibinfo {author}
  {\bibfnamefont {X.}~\bibnamefont {Dai}}, \ and\ \bibinfo {author}
  {\bibfnamefont {Z.}~\bibnamefont {Fang}},\ }\href@noop {} {\bibfield
  {journal} {\bibinfo  {journal} {Physical Review B},\ }\textbf {\bibinfo
  {volume} {88}},\ \bibinfo {pages} {125427} (\bibinfo {year}
  {2013})}\BibitemShut {NoStop}%
\bibitem [{\citenamefont {Wang}\ \emph {et~al.}(2012)\citenamefont {Wang},
  \citenamefont {Sun}, \citenamefont {Chen}, \citenamefont {Franchini},
  \citenamefont {Xu}, \citenamefont {Weng}, \citenamefont {Dai},\ and\
  \citenamefont {Fang}}]{wang2012dirac}%
  \BibitemOpen
  \bibfield  {author} {\bibinfo {author} {\bibfnamefont {Z.}~\bibnamefont
  {Wang}}, \bibinfo {author} {\bibfnamefont {Y.}~\bibnamefont {Sun}}, \bibinfo
  {author} {\bibfnamefont {X.-Q.}\ \bibnamefont {Chen}}, \bibinfo {author}
  {\bibfnamefont {C.}~\bibnamefont {Franchini}}, \bibinfo {author}
  {\bibfnamefont {G.}~\bibnamefont {Xu}}, \bibinfo {author} {\bibfnamefont
  {H.}~\bibnamefont {Weng}}, \bibinfo {author} {\bibfnamefont {X.}~\bibnamefont
  {Dai}}, \ and\ \bibinfo {author} {\bibfnamefont {Z.}~\bibnamefont {Fang}},\
  }\href@noop {} {\bibfield  {journal} {\bibinfo  {journal} {Physical Review
  B},\ }\textbf {\bibinfo {volume} {85}},\ \bibinfo {pages} {195320} (\bibinfo
  {year} {2012})}\BibitemShut {NoStop}%
\bibitem [{\citenamefont {Borisenko}\ \emph {et~al.}(2014)\citenamefont
  {Borisenko}, \citenamefont {Gibson}, \citenamefont {Evtushinsky},
  \citenamefont {Zabolotnyy}, \citenamefont {B\"uchner},\ and\ \citenamefont
  {Cava}}]{PhysRevLett.113.027603}%
  \BibitemOpen
  \bibfield  {author} {\bibinfo {author} {\bibfnamefont {S.}~\bibnamefont
  {Borisenko}}, \bibinfo {author} {\bibfnamefont {Q.}~\bibnamefont {Gibson}},
  \bibinfo {author} {\bibfnamefont {D.}~\bibnamefont {Evtushinsky}}, \bibinfo
  {author} {\bibfnamefont {V.}~\bibnamefont {Zabolotnyy}}, \bibinfo {author}
  {\bibfnamefont {B.}~\bibnamefont {B\"uchner}}, \ and\ \bibinfo {author}
  {\bibfnamefont {R.~J.}\ \bibnamefont {Cava}},\ }\Doi
  {10.1103/PhysRevLett.113.027603} {\bibfield  {journal} {\bibinfo  {journal}
  {Phys. Rev. Lett.},\ }\textbf {\bibinfo {volume} {113}},\ \bibinfo {pages}
  {027603} (\bibinfo {year} {2014})}\BibitemShut {NoStop}%
\bibitem [{\citenamefont {Neupane}\ \emph {et~al.}(2014)\citenamefont
  {Neupane}, \citenamefont {Xu}, \citenamefont {Sankar}, \citenamefont
  {Alidoust}, \citenamefont {Bian}, \citenamefont {Liu}, \citenamefont
  {Belopolski}, \citenamefont {Chang}, \citenamefont {Jeng}, \citenamefont
  {Lin} \emph {et~al.}}]{neupane2014observation}%
  \BibitemOpen
  \bibfield  {author} {\bibinfo {author} {\bibfnamefont {M.}~\bibnamefont
  {Neupane}}, \bibinfo {author} {\bibfnamefont {S.-Y.}\ \bibnamefont {Xu}},
  \bibinfo {author} {\bibfnamefont {R.}~\bibnamefont {Sankar}}, \bibinfo
  {author} {\bibfnamefont {N.}~\bibnamefont {Alidoust}}, \bibinfo {author}
  {\bibfnamefont {G.}~\bibnamefont {Bian}}, \bibinfo {author} {\bibfnamefont
  {C.}~\bibnamefont {Liu}}, \bibinfo {author} {\bibfnamefont {I.}~\bibnamefont
  {Belopolski}}, \bibinfo {author} {\bibfnamefont {T.-R.}\ \bibnamefont
  {Chang}}, \bibinfo {author} {\bibfnamefont {H.-T.}\ \bibnamefont {Jeng}},
  \bibinfo {author} {\bibfnamefont {H.}~\bibnamefont {Lin}},  \emph {et~al.},\
  }\href@noop {} {\bibfield  {journal} {\bibinfo  {journal} {Nature
  communications},\ }\textbf {\bibinfo {volume} {5}} (\bibinfo {year}
  {2014})}\BibitemShut {NoStop}%
\bibitem [{\citenamefont {Liu}\ \emph {et~al.}(2014){\natexlab{a}}\citenamefont
  {Liu}, \citenamefont {Zhou}, \citenamefont {Zhang}, \citenamefont {Wang},
  \citenamefont {Weng}, \citenamefont {Prabhakaran}, \citenamefont {Mo},
  \citenamefont {Shen}, \citenamefont {Fang}, \citenamefont {Dai} \emph
  {et~al.}}]{liu2014discovery}%
  \BibitemOpen
  \bibfield  {author} {\bibinfo {author} {\bibfnamefont {Z.}~\bibnamefont
  {Liu}}, \bibinfo {author} {\bibfnamefont {B.}~\bibnamefont {Zhou}}, \bibinfo
  {author} {\bibfnamefont {Y.}~\bibnamefont {Zhang}}, \bibinfo {author}
  {\bibfnamefont {Z.}~\bibnamefont {Wang}}, \bibinfo {author} {\bibfnamefont
  {H.}~\bibnamefont {Weng}}, \bibinfo {author} {\bibfnamefont {D.}~\bibnamefont
  {Prabhakaran}}, \bibinfo {author} {\bibfnamefont {S.-K.}\ \bibnamefont {Mo}},
  \bibinfo {author} {\bibfnamefont {Z.}~\bibnamefont {Shen}}, \bibinfo {author}
  {\bibfnamefont {Z.}~\bibnamefont {Fang}}, \bibinfo {author} {\bibfnamefont
  {X.}~\bibnamefont {Dai}},  \emph {et~al.},\ }\href@noop {} {\bibfield
  {journal} {\bibinfo  {journal} {Science},\ }\textbf {\bibinfo {volume}
  {343}},\ \bibinfo {pages} {864} (\bibinfo {year}
  {2014}{\natexlab{a}})}\BibitemShut {NoStop}%
\bibitem [{\citenamefont {Liu}\ \emph {et~al.}(2014){\natexlab{b}}\citenamefont
  {Liu}, \citenamefont {Jiang}, \citenamefont {Zhou}, \citenamefont {Wang},
  \citenamefont {Zhang}, \citenamefont {Weng}, \citenamefont {Prabhakaran},
  \citenamefont {Mo}, \citenamefont {Peng}, \citenamefont {Dudin} \emph
  {et~al.}}]{liu2014stable}%
  \BibitemOpen
  \bibfield  {author} {\bibinfo {author} {\bibfnamefont {Z.}~\bibnamefont
  {Liu}}, \bibinfo {author} {\bibfnamefont {J.}~\bibnamefont {Jiang}}, \bibinfo
  {author} {\bibfnamefont {B.}~\bibnamefont {Zhou}}, \bibinfo {author}
  {\bibfnamefont {Z.}~\bibnamefont {Wang}}, \bibinfo {author} {\bibfnamefont
  {Y.}~\bibnamefont {Zhang}}, \bibinfo {author} {\bibfnamefont
  {H.}~\bibnamefont {Weng}}, \bibinfo {author} {\bibfnamefont {D.}~\bibnamefont
  {Prabhakaran}}, \bibinfo {author} {\bibfnamefont {S.}~\bibnamefont {Mo}},
  \bibinfo {author} {\bibfnamefont {H.}~\bibnamefont {Peng}}, \bibinfo {author}
  {\bibfnamefont {P.}~\bibnamefont {Dudin}},  \emph {et~al.},\ }\href@noop {}
  {\bibfield  {journal} {\bibinfo  {journal} {Nature materials}} (\bibinfo
  {year} {2014}{\natexlab{b}})}\BibitemShut {NoStop}%
\bibitem [{\citenamefont {Jeon}\ \emph {et~al.}(2014)\citenamefont {Jeon},
  \citenamefont {Zhou}, \citenamefont {Gyenis}, \citenamefont {Feldman},
  \citenamefont {Kimchi}, \citenamefont {Potter}, \citenamefont {Gibson},
  \citenamefont {Cava}, \citenamefont {Vishwanath},\ and\ \citenamefont
  {Yazdani}}]{jeon2014landau}%
  \BibitemOpen
  \bibfield  {author} {\bibinfo {author} {\bibfnamefont {S.}~\bibnamefont
  {Jeon}}, \bibinfo {author} {\bibfnamefont {B.~B.}\ \bibnamefont {Zhou}},
  \bibinfo {author} {\bibfnamefont {A.}~\bibnamefont {Gyenis}}, \bibinfo
  {author} {\bibfnamefont {B.~E.}\ \bibnamefont {Feldman}}, \bibinfo {author}
  {\bibfnamefont {I.}~\bibnamefont {Kimchi}}, \bibinfo {author} {\bibfnamefont
  {A.~C.}\ \bibnamefont {Potter}}, \bibinfo {author} {\bibfnamefont {Q.~D.}\
  \bibnamefont {Gibson}}, \bibinfo {author} {\bibfnamefont {R.~J.}\
  \bibnamefont {Cava}}, \bibinfo {author} {\bibfnamefont {A.}~\bibnamefont
  {Vishwanath}}, \ and\ \bibinfo {author} {\bibfnamefont {A.}~\bibnamefont
  {Yazdani}},\ }\href@noop {} {\bibfield  {journal} {\bibinfo  {journal} {arXiv
  preprint arXiv:1403.3446}} (\bibinfo {year} {2014})}\BibitemShut {NoStop}%
\bibitem [{\citenamefont {Feng}\ \emph {et~al.}(2014)\citenamefont {Feng},
  \citenamefont {Pang}, \citenamefont {Wu}, \citenamefont {Wang}, \citenamefont
  {Weng}, \citenamefont {Li}, \citenamefont {Dai}, \citenamefont {Fang},
  \citenamefont {Shi},\ and\ \citenamefont {Lu}}]{feng2014large}%
  \BibitemOpen
  \bibfield  {author} {\bibinfo {author} {\bibfnamefont {J.}~\bibnamefont
  {Feng}}, \bibinfo {author} {\bibfnamefont {Y.}~\bibnamefont {Pang}}, \bibinfo
  {author} {\bibfnamefont {D.}~\bibnamefont {Wu}}, \bibinfo {author}
  {\bibfnamefont {Z.}~\bibnamefont {Wang}}, \bibinfo {author} {\bibfnamefont
  {H.}~\bibnamefont {Weng}}, \bibinfo {author} {\bibfnamefont {J.}~\bibnamefont
  {Li}}, \bibinfo {author} {\bibfnamefont {X.}~\bibnamefont {Dai}}, \bibinfo
  {author} {\bibfnamefont {Z.}~\bibnamefont {Fang}}, \bibinfo {author}
  {\bibfnamefont {Y.}~\bibnamefont {Shi}}, \ and\ \bibinfo {author}
  {\bibfnamefont {L.}~\bibnamefont {Lu}},\ }\href@noop {} {\bibfield  {journal}
  {\bibinfo  {journal} {arXiv preprint arXiv:1405.6611}} (\bibinfo {year}
  {2014})}\BibitemShut {NoStop}%
\bibitem [{\citenamefont {Liang}\ \emph {et~al.}(2014)\citenamefont {Liang},
  \citenamefont {Gibson}, \citenamefont {Ali}, \citenamefont {Liu},
  \citenamefont {Cava},\ and\ \citenamefont {Ong}}]{liang2014ultrahigh}%
  \BibitemOpen
  \bibfield  {author} {\bibinfo {author} {\bibfnamefont {T.}~\bibnamefont
  {Liang}}, \bibinfo {author} {\bibfnamefont {Q.}~\bibnamefont {Gibson}},
  \bibinfo {author} {\bibfnamefont {M.~N.}\ \bibnamefont {Ali}}, \bibinfo
  {author} {\bibfnamefont {M.}~\bibnamefont {Liu}}, \bibinfo {author}
  {\bibfnamefont {R.}~\bibnamefont {Cava}}, \ and\ \bibinfo {author}
  {\bibfnamefont {N.}~\bibnamefont {Ong}},\ }\href@noop {} {\bibfield
  {journal} {\bibinfo  {journal} {arXiv preprint arXiv:1404.7794}} (\bibinfo
  {year} {2014})}\BibitemShut {NoStop}%
\bibitem [{\citenamefont {Xu}\ \emph {et~al.}(2013)\citenamefont {Xu},
  \citenamefont {Liu}, \citenamefont {Kushwaha}, \citenamefont {Chang},
  \citenamefont {Krizan}, \citenamefont {Sankar}, \citenamefont {Polley},
  \citenamefont {Adell}, \citenamefont {Balasubramanian}, \citenamefont
  {Miyamoto} \emph {et~al.}}]{xu2013observation}%
  \BibitemOpen
  \bibfield  {author} {\bibinfo {author} {\bibfnamefont {S.-Y.}\ \bibnamefont
  {Xu}}, \bibinfo {author} {\bibfnamefont {C.}~\bibnamefont {Liu}}, \bibinfo
  {author} {\bibfnamefont {S.}~\bibnamefont {Kushwaha}}, \bibinfo {author}
  {\bibfnamefont {T.-R.}\ \bibnamefont {Chang}}, \bibinfo {author}
  {\bibfnamefont {J.}~\bibnamefont {Krizan}}, \bibinfo {author} {\bibfnamefont
  {R.}~\bibnamefont {Sankar}}, \bibinfo {author} {\bibfnamefont
  {C.}~\bibnamefont {Polley}}, \bibinfo {author} {\bibfnamefont
  {J.}~\bibnamefont {Adell}}, \bibinfo {author} {\bibfnamefont
  {T.}~\bibnamefont {Balasubramanian}}, \bibinfo {author} {\bibfnamefont
  {K.}~\bibnamefont {Miyamoto}},  \emph {et~al.},\ }\href@noop {} {\bibfield
  {journal} {\bibinfo  {journal} {arXiv preprint arXiv:1312.7624}} (\bibinfo
  {year} {2013})}\BibitemShut {NoStop}%
\bibitem [{\citenamefont {Zhu}\ and\ \citenamefont
  {Hoffman}(2014)}]{zhu2014condensed}%
  \BibitemOpen
  \bibfield  {author} {\bibinfo {author} {\bibfnamefont {Z.}~\bibnamefont
  {Zhu}}\ and\ \bibinfo {author} {\bibfnamefont {J.~E.}\ \bibnamefont
  {Hoffman}},\ }\href@noop {} {\bibfield  {journal} {\bibinfo  {journal}
  {Nature},\ }\textbf {\bibinfo {volume} {513}},\ \bibinfo {pages} {319}
  (\bibinfo {year} {2014})}\BibitemShut {NoStop}%
\bibitem [{\citenamefont {Park}\ \emph {et~al.}(2011)\citenamefont {Park},
  \citenamefont {Lee}, \citenamefont {Wolff-Fabris}, \citenamefont {Koh},
  \citenamefont {Eom}, \citenamefont {Kim}, \citenamefont {Farhan},
  \citenamefont {Jo}, \citenamefont {Kim}, \citenamefont {Shim} \emph
  {et~al.}}]{park2011anisotropic}%
  \BibitemOpen
  \bibfield  {author} {\bibinfo {author} {\bibfnamefont {J.}~\bibnamefont
  {Park}}, \bibinfo {author} {\bibfnamefont {G.}~\bibnamefont {Lee}}, \bibinfo
  {author} {\bibfnamefont {F.}~\bibnamefont {Wolff-Fabris}}, \bibinfo {author}
  {\bibfnamefont {Y.}~\bibnamefont {Koh}}, \bibinfo {author} {\bibfnamefont
  {M.}~\bibnamefont {Eom}}, \bibinfo {author} {\bibfnamefont {Y.}~\bibnamefont
  {Kim}}, \bibinfo {author} {\bibfnamefont {M.}~\bibnamefont {Farhan}},
  \bibinfo {author} {\bibfnamefont {Y.}~\bibnamefont {Jo}}, \bibinfo {author}
  {\bibfnamefont {C.}~\bibnamefont {Kim}}, \bibinfo {author} {\bibfnamefont
  {J.}~\bibnamefont {Shim}},  \emph {et~al.},\ }\href@noop {} {\bibfield
  {journal} {\bibinfo  {journal} {Physical review letters},\ }\textbf {\bibinfo
  {volume} {107}},\ \bibinfo {pages} {126402} (\bibinfo {year}
  {2011})}\BibitemShut {NoStop}%
\bibitem [{\citenamefont {Sun}\ \emph {et~al.}(2010)\citenamefont {Sun},
  \citenamefont {Chen}, \citenamefont {Yunoki}, \citenamefont {Li},\ and\
  \citenamefont {Li}}]{sun2010new}%
  \BibitemOpen
  \bibfield  {author} {\bibinfo {author} {\bibfnamefont {Y.}~\bibnamefont
  {Sun}}, \bibinfo {author} {\bibfnamefont {X.-Q.}\ \bibnamefont {Chen}},
  \bibinfo {author} {\bibfnamefont {S.}~\bibnamefont {Yunoki}}, \bibinfo
  {author} {\bibfnamefont {D.}~\bibnamefont {Li}}, \ and\ \bibinfo {author}
  {\bibfnamefont {Y.}~\bibnamefont {Li}},\ }\href@noop {} {\bibfield  {journal}
  {\bibinfo  {journal} {Physical review letters},\ }\textbf {\bibinfo {volume}
  {105}},\ \bibinfo {pages} {216406} (\bibinfo {year} {2010})}\BibitemShut
  {NoStop}%
\bibitem [{\citenamefont {Hsieh}\ \emph {et~al.}(2014)\citenamefont {Hsieh},
  \citenamefont {Liu},\ and\ \citenamefont {Fu}}]{hsieh2014topological2}%
  \BibitemOpen
  \bibfield  {author} {\bibinfo {author} {\bibfnamefont {T.~H.}\ \bibnamefont
  {Hsieh}}, \bibinfo {author} {\bibfnamefont {J.}~\bibnamefont {Liu}}, \ and\
  \bibinfo {author} {\bibfnamefont {L.}~\bibnamefont {Fu}},\ }\href@noop {}
  {\bibfield  {journal} {\bibinfo  {journal} {Physical Review B},\ }\textbf
  {\bibinfo {volume} {90}},\ \bibinfo {pages} {081112} (\bibinfo {year}
  {2014})}\BibitemShut {NoStop}%
\bibitem [{\citenamefont {Harrison}\ and\ \citenamefont
  {Sebastian}(2009)}]{harrison2009dirac}%
  \BibitemOpen
  \bibfield  {author} {\bibinfo {author} {\bibfnamefont {N.}~\bibnamefont
  {Harrison}}\ and\ \bibinfo {author} {\bibfnamefont {S.}~\bibnamefont
  {Sebastian}},\ }\href@noop {} {\bibfield  {journal} {\bibinfo  {journal}
  {Physical Review B},\ }\textbf {\bibinfo {volume} {80}},\ \bibinfo {pages}
  {224512} (\bibinfo {year} {2009})}\BibitemShut {NoStop}%
\bibitem [{\citenamefont {Morinari}\ \emph {et~al.}(2010)\citenamefont
  {Morinari}, \citenamefont {Kaneshita},\ and\ \citenamefont
  {Tohyama}}]{morinari2010topological}%
  \BibitemOpen
  \bibfield  {author} {\bibinfo {author} {\bibfnamefont {T.}~\bibnamefont
  {Morinari}}, \bibinfo {author} {\bibfnamefont {E.}~\bibnamefont {Kaneshita}},
  \ and\ \bibinfo {author} {\bibfnamefont {T.}~\bibnamefont {Tohyama}},\
  }\href@noop {} {\bibfield  {journal} {\bibinfo  {journal} {Physical review
  letters},\ }\textbf {\bibinfo {volume} {105}},\ \bibinfo {pages} {037203}
  (\bibinfo {year} {2010})}\BibitemShut {NoStop}%
\bibitem [{\citenamefont {Richard}\ \emph {et~al.}(2010)\citenamefont
  {Richard}, \citenamefont {Nakayama}, \citenamefont {Sato}, \citenamefont
  {Neupane}, \citenamefont {Xu}, \citenamefont {Bowen}, \citenamefont {Chen},
  \citenamefont {Luo}, \citenamefont {Wang}, \citenamefont {Dai} \emph
  {et~al.}}]{richard2010observation}%
  \BibitemOpen
  \bibfield  {author} {\bibinfo {author} {\bibfnamefont {P.}~\bibnamefont
  {Richard}}, \bibinfo {author} {\bibfnamefont {K.}~\bibnamefont {Nakayama}},
  \bibinfo {author} {\bibfnamefont {T.}~\bibnamefont {Sato}}, \bibinfo {author}
  {\bibfnamefont {M.}~\bibnamefont {Neupane}}, \bibinfo {author} {\bibfnamefont
  {Y.-M.}\ \bibnamefont {Xu}}, \bibinfo {author} {\bibfnamefont
  {J.}~\bibnamefont {Bowen}}, \bibinfo {author} {\bibfnamefont
  {G.}~\bibnamefont {Chen}}, \bibinfo {author} {\bibfnamefont {J.}~\bibnamefont
  {Luo}}, \bibinfo {author} {\bibfnamefont {N.}~\bibnamefont {Wang}}, \bibinfo
  {author} {\bibfnamefont {X.}~\bibnamefont {Dai}},  \emph {et~al.},\
  }\href@noop {} {\bibfield  {journal} {\bibinfo  {journal} {Physical review
  letters},\ }\textbf {\bibinfo {volume} {104}},\ \bibinfo {pages} {137001}
  (\bibinfo {year} {2010})}\BibitemShut {NoStop}%
\bibitem [{\citenamefont {Elliott}(1954)}]{PhysRev.96.280}%
  \BibitemOpen
  \bibfield  {author} {\bibinfo {author} {\bibfnamefont {R.~J.}\ \bibnamefont
  {Elliott}},\ }\Doi {10.1103/PhysRev.96.280} {\bibfield  {journal} {\bibinfo
  {journal} {Phys. Rev.},\ }\textbf {\bibinfo {volume} {96}},\ \bibinfo {pages}
  {280} (\bibinfo {year} {1954})}\BibitemShut {NoStop}%
\bibitem [{\citenamefont {Dresselhaus}(1955)}]{PhysRev.100.580}%
  \BibitemOpen
  \bibfield  {author} {\bibinfo {author} {\bibfnamefont {G.}~\bibnamefont
  {Dresselhaus}},\ }\Doi {10.1103/PhysRev.100.580} {\bibfield  {journal}
  {\bibinfo  {journal} {Phys. Rev.},\ }\textbf {\bibinfo {volume} {100}},\
  \bibinfo {pages} {580} (\bibinfo {year} {1955})}\BibitemShut {NoStop}%
\bibitem [{\citenamefont {Koster}(1963)}]{koster1963properties}%
  \BibitemOpen
  \bibfield  {author} {\bibinfo {author} {\bibfnamefont {G.}~\bibnamefont
  {Koster}},\ }\href@noop {} {\emph {\bibinfo {title} {Properties of the
  thirty-two point groups}}},\ Massachusetts institute of technology press
  research monograph\ (\bibinfo  {publisher} {M.I.T. Press},\ \bibinfo {year}
  {1963})\BibitemShut {NoStop}%
\bibitem [{\citenamefont {Xie}\ \emph {et~al.}(2014)\citenamefont {Xie},
  \citenamefont {Schoop}, \citenamefont {Medvedev}, \citenamefont {Felser},\
  and\ \citenamefont {Cava}}]{xie2014pressure}%
  \BibitemOpen
  \bibfield  {author} {\bibinfo {author} {\bibfnamefont {L.~S.}\ \bibnamefont
  {Xie}}, \bibinfo {author} {\bibfnamefont {L.~M.}\ \bibnamefont {Schoop}},
  \bibinfo {author} {\bibfnamefont {S.~A.}\ \bibnamefont {Medvedev}}, \bibinfo
  {author} {\bibfnamefont {C.}~\bibnamefont {Felser}}, \ and\ \bibinfo {author}
  {\bibfnamefont {R.}~\bibnamefont {Cava}},\ }\href@noop {} {\bibfield
  {journal} {\bibinfo  {journal} {Solid State Sciences},\ }\textbf {\bibinfo
  {volume} {30}},\ \bibinfo {pages} {6} (\bibinfo {year} {2014})}\BibitemShut
  {NoStop}%
\bibitem [{\citenamefont {Kang}\ and\ \citenamefont
  {Miller}(2002)}]{kang2002intermetallic}%
  \BibitemOpen
  \bibfield  {author} {\bibinfo {author} {\bibfnamefont {S.~K.}\ \bibnamefont
  {Kang}}\ and\ \bibinfo {author} {\bibfnamefont {G.~J.}\ \bibnamefont
  {Miller}},\ }\href@noop {} {\bibfield  {journal} {\bibinfo  {journal} {Acta
  Crystallographica Section E: Structure Reports Online},\ }\textbf {\bibinfo
  {volume} {58}},\ \bibinfo {pages} {i21} (\bibinfo {year} {2002})}\BibitemShut
  {NoStop}%
\bibitem [{\citenamefont {Chadov}\ \emph {et~al.}(2010)\citenamefont {Chadov},
  \citenamefont {Qi}, \citenamefont {K{\"u}bler}, \citenamefont {Fecher},
  \citenamefont {Felser},\ and\ \citenamefont {Zhang}}]{chadov2010tunable}%
  \BibitemOpen
  \bibfield  {author} {\bibinfo {author} {\bibfnamefont {S.}~\bibnamefont
  {Chadov}}, \bibinfo {author} {\bibfnamefont {X.}~\bibnamefont {Qi}}, \bibinfo
  {author} {\bibfnamefont {J.}~\bibnamefont {K{\"u}bler}}, \bibinfo {author}
  {\bibfnamefont {G.~H.}\ \bibnamefont {Fecher}}, \bibinfo {author}
  {\bibfnamefont {C.}~\bibnamefont {Felser}}, \ and\ \bibinfo {author}
  {\bibfnamefont {S.~C.}\ \bibnamefont {Zhang}},\ }\href@noop {} {\bibfield
  {journal} {\bibinfo  {journal} {Nature materials},\ }\textbf {\bibinfo
  {volume} {9}},\ \bibinfo {pages} {541} (\bibinfo {year} {2010})}\BibitemShut
  {NoStop}%
\bibitem [{\citenamefont {Brese}\ and\ \citenamefont {von
  Schnering}(1994)}]{brese1994bonding}%
  \BibitemOpen
  \bibfield  {author} {\bibinfo {author} {\bibfnamefont {N.~E.}\ \bibnamefont
  {Brese}}\ and\ \bibinfo {author} {\bibfnamefont {H.~G.}\ \bibnamefont {von
  Schnering}},\ }\href@noop {} {\bibfield  {journal} {\bibinfo  {journal}
  {Zeitschrift f{\"u}r anorganische und allgemeine Chemie},\ }\textbf {\bibinfo
  {volume} {620}},\ \bibinfo {pages} {393} (\bibinfo {year}
  {1994})}\BibitemShut {NoStop}%
\bibitem [{\citenamefont {Villars}\ \emph {et~al.}(2008)\citenamefont
  {Villars}, \citenamefont {Cenzual}, \citenamefont {Daams}, \citenamefont
  {Gladyshevskii}, \citenamefont {Shcherban}, \citenamefont {Dubenskyy},
  \citenamefont {Melnichenko-Koblyuk}, \citenamefont {Pavlyuk}, \citenamefont
  {Savysyuk}, \citenamefont {Stoyko} \emph {et~al.}}]{villars2008srsn2as2}%
  \BibitemOpen
  \bibfield  {author} {\bibinfo {author} {\bibfnamefont {P.}~\bibnamefont
  {Villars}}, \bibinfo {author} {\bibfnamefont {K.}~\bibnamefont {Cenzual}},
  \bibinfo {author} {\bibfnamefont {J.}~\bibnamefont {Daams}}, \bibinfo
  {author} {\bibfnamefont {R.}~\bibnamefont {Gladyshevskii}}, \bibinfo {author}
  {\bibfnamefont {O.}~\bibnamefont {Shcherban}}, \bibinfo {author}
  {\bibfnamefont {V.}~\bibnamefont {Dubenskyy}}, \bibinfo {author}
  {\bibfnamefont {N.}~\bibnamefont {Melnichenko-Koblyuk}}, \bibinfo {author}
  {\bibfnamefont {O.}~\bibnamefont {Pavlyuk}}, \bibinfo {author} {\bibfnamefont
  {I.}~\bibnamefont {Savysyuk}}, \bibinfo {author} {\bibfnamefont
  {S.}~\bibnamefont {Stoyko}},  \emph {et~al.},\ }in\ \href@noop {} {\emph
  {\bibinfo {booktitle} {Structure Types. Part 6: Space Groups (166) R-3m-(160)
  R3m}}}\ (\bibinfo  {publisher} {Springer},\ \bibinfo {year} {2008})\ pp.\
  \bibinfo {pages} {681--681}\BibitemShut {NoStop}%
\bibitem [{\citenamefont {Buzea}\ and\ \citenamefont
  {Yamashita}(2001)}]{buzea2001review}%
  \BibitemOpen
  \bibfield  {author} {\bibinfo {author} {\bibfnamefont {C.}~\bibnamefont
  {Buzea}}\ and\ \bibinfo {author} {\bibfnamefont {T.}~\bibnamefont
  {Yamashita}},\ }\href@noop {} {\bibfield  {journal} {\bibinfo  {journal}
  {Superconductor Science and Technology},\ }\textbf {\bibinfo {volume} {14}},\
  \bibinfo {pages} {R115} (\bibinfo {year} {2001})}\BibitemShut {NoStop}%
\bibitem [{\citenamefont {Iandelli}(1955)}]{iandelli1955structure}%
  \BibitemOpen
  \bibfield  {author} {\bibinfo {author} {\bibfnamefont {A.}~\bibnamefont
  {Iandelli}},\ }\href@noop {} {\bibfield  {journal} {\bibinfo  {journal} {Atti
  Accad. Naz. Lincei, Cl. Sci. Fis. Mat. Nat},\ }\textbf {\bibinfo {volume}
  {19}},\ \bibinfo {pages} {39} (\bibinfo {year} {1955})}\BibitemShut {NoStop}%
\bibitem [{\citenamefont {Rasche}\ \emph {et~al.}(2013)\citenamefont {Rasche},
  \citenamefont {Isaeva}, \citenamefont {Gerisch}, \citenamefont {Kaiser},
  \citenamefont {Van~den Broek}, \citenamefont {Koch}, \citenamefont {Kaiser},\
  and\ \citenamefont {Ruck}}]{rasche2013crystal}%
  \BibitemOpen
  \bibfield  {author} {\bibinfo {author} {\bibfnamefont {B.}~\bibnamefont
  {Rasche}}, \bibinfo {author} {\bibfnamefont {A.}~\bibnamefont {Isaeva}},
  \bibinfo {author} {\bibfnamefont {A.}~\bibnamefont {Gerisch}}, \bibinfo
  {author} {\bibfnamefont {M.}~\bibnamefont {Kaiser}}, \bibinfo {author}
  {\bibfnamefont {W.}~\bibnamefont {Van~den Broek}}, \bibinfo {author}
  {\bibfnamefont {C.~T.}\ \bibnamefont {Koch}}, \bibinfo {author}
  {\bibfnamefont {U.}~\bibnamefont {Kaiser}}, \ and\ \bibinfo {author}
  {\bibfnamefont {M.}~\bibnamefont {Ruck}},\ }\href@noop {} {\bibfield
  {journal} {\bibinfo  {journal} {Chemistry of Materials},\ }\textbf {\bibinfo
  {volume} {25}},\ \bibinfo {pages} {2359} (\bibinfo {year}
  {2013})}\BibitemShut {NoStop}%
\bibitem [{\citenamefont {Song}\ \emph {et~al.}(2014)\citenamefont {Song},
  \citenamefont {Liu}, \citenamefont {Yang}, \citenamefont {Han}, \citenamefont
  {Ye}, \citenamefont {Fu}, \citenamefont {Yang}, \citenamefont {Niu},
  \citenamefont {Lu},\ and\ \citenamefont {Yao}}]{song2014quantum}%
  \BibitemOpen
  \bibfield  {author} {\bibinfo {author} {\bibfnamefont {Z.}~\bibnamefont
  {Song}}, \bibinfo {author} {\bibfnamefont {C.-C.}\ \bibnamefont {Liu}},
  \bibinfo {author} {\bibfnamefont {J.}~\bibnamefont {Yang}}, \bibinfo {author}
  {\bibfnamefont {J.}~\bibnamefont {Han}}, \bibinfo {author} {\bibfnamefont
  {M.}~\bibnamefont {Ye}}, \bibinfo {author} {\bibfnamefont {B.}~\bibnamefont
  {Fu}}, \bibinfo {author} {\bibfnamefont {Y.}~\bibnamefont {Yang}}, \bibinfo
  {author} {\bibfnamefont {Q.}~\bibnamefont {Niu}}, \bibinfo {author}
  {\bibfnamefont {J.}~\bibnamefont {Lu}}, \ and\ \bibinfo {author}
  {\bibfnamefont {Y.}~\bibnamefont {Yao}},\ }\href@noop {} {\bibfield
  {journal} {\bibinfo  {journal} {arXiv preprint arXiv:1402.2399}} (\bibinfo
  {year} {2014})}\BibitemShut {NoStop}%
\bibitem [{\citenamefont {Ali}\ \emph {et~al.}(2014)\citenamefont {Ali},
  \citenamefont {Gibson}, \citenamefont {Klimczuk},\ and\ \citenamefont
  {Cava}}]{PhysRevB.89.020505}%
  \BibitemOpen
  \bibfield  {author} {\bibinfo {author} {\bibfnamefont {M.~N.}\ \bibnamefont
  {Ali}}, \bibinfo {author} {\bibfnamefont {Q.~D.}\ \bibnamefont {Gibson}},
  \bibinfo {author} {\bibfnamefont {T.}~\bibnamefont {Klimczuk}}, \ and\
  \bibinfo {author} {\bibfnamefont {R.~J.}\ \bibnamefont {Cava}},\ }\Doi
  {10.1103/PhysRevB.89.020505} {\bibfield  {journal} {\bibinfo  {journal}
  {Phys. Rev. B},\ }\textbf {\bibinfo {volume} {89}},\ \bibinfo {pages}
  {020505} (\bibinfo {year} {2014})}\BibitemShut {NoStop}%
\bibitem [{\citenamefont {Karpov}\ \emph {et~al.}(2004)\citenamefont {Karpov},
  \citenamefont {Nuss}, \citenamefont {Wedig},\ and\ \citenamefont
  {Jansen}}]{karpov2004covalently}%
  \BibitemOpen
  \bibfield  {author} {\bibinfo {author} {\bibfnamefont {A.}~\bibnamefont
  {Karpov}}, \bibinfo {author} {\bibfnamefont {J.}~\bibnamefont {Nuss}},
  \bibinfo {author} {\bibfnamefont {U.}~\bibnamefont {Wedig}}, \ and\ \bibinfo
  {author} {\bibfnamefont {M.}~\bibnamefont {Jansen}},\ }\href@noop {}
  {\bibfield  {journal} {\bibinfo  {journal} {Journal of the American Chemical
  Society},\ }\textbf {\bibinfo {volume} {126}},\ \bibinfo {pages} {14123}
  (\bibinfo {year} {2004})}\BibitemShut {NoStop}%
\bibitem [{\citenamefont {Bronger}\ \emph {et~al.}(1975)\citenamefont
  {Bronger}, \citenamefont {Nacken},\ and\ \citenamefont
  {Ploog}}]{bronger1975synthesis}%
  \BibitemOpen
  \bibfield  {author} {\bibinfo {author} {\bibfnamefont {W.}~\bibnamefont
  {Bronger}}, \bibinfo {author} {\bibfnamefont {B.}~\bibnamefont {Nacken}}, \
  and\ \bibinfo {author} {\bibfnamefont {K.}~\bibnamefont {Ploog}},\
  }\href@noop {} {\bibfield  {journal} {\bibinfo  {journal} {J. Less-Common
  Met.},\ }\textbf {\bibinfo {volume} {43}},\ \bibinfo {pages} {143} (\bibinfo
  {year} {1975})}\BibitemShut {NoStop}%
\bibitem [{\citenamefont {Zak}(2001)}]{zak2001topologically}%
  \BibitemOpen
  \bibfield  {author} {\bibinfo {author} {\bibfnamefont {J.}~\bibnamefont
  {Zak}},\ }\href@noop {} {\emph {\bibinfo {title} {Topologically Unavoidable
  Degeneracies in Band Structure of Solids}}}\ (\bibinfo {year} {2001})\ p.\
  \bibinfo {pages} {222}\BibitemShut {NoStop}%
\bibitem [{\citenamefont {Michel}\ and\ \citenamefont
  {Zak}(1999)}]{PhysRevB.59.5998}%
  \BibitemOpen
  \bibfield  {author} {\bibinfo {author} {\bibfnamefont {L.}~\bibnamefont
  {Michel}}\ and\ \bibinfo {author} {\bibfnamefont {J.}~\bibnamefont {Zak}},\
  }\Doi {10.1103/PhysRevB.59.5998} {\bibfield  {journal} {\bibinfo  {journal}
  {Phys. Rev. B},\ }\textbf {\bibinfo {volume} {59}},\ \bibinfo {pages} {5998}
  (\bibinfo {year} {1999})}\BibitemShut {NoStop}%
\bibitem [{\citenamefont {Parameswaran}\ \emph {et~al.}(2013)\citenamefont
  {Parameswaran}, \citenamefont {Turner}, \citenamefont {Arovas},\ and\
  \citenamefont {Vishwanath}}]{parameswaran2013topological}%
  \BibitemOpen
  \bibfield  {author} {\bibinfo {author} {\bibfnamefont {S.~A.}\ \bibnamefont
  {Parameswaran}}, \bibinfo {author} {\bibfnamefont {A.~M.}\ \bibnamefont
  {Turner}}, \bibinfo {author} {\bibfnamefont {D.~P.}\ \bibnamefont {Arovas}},
  \ and\ \bibinfo {author} {\bibfnamefont {A.}~\bibnamefont {Vishwanath}},\
  }\href@noop {} {\bibfield  {journal} {\bibinfo  {journal} {Nature Physics},\
  }\textbf {\bibinfo {volume} {9}},\ \bibinfo {pages} {299} (\bibinfo {year}
  {2013})}\BibitemShut {NoStop}%
\bibitem [{\citenamefont {Young}\ \emph {et~al.}(2012)\citenamefont {Young},
  \citenamefont {Zaheer}, \citenamefont {Teo}, \citenamefont {Kane},
  \citenamefont {Mele},\ and\ \citenamefont {Rappe}}]{young2012dirac}%
  \BibitemOpen
  \bibfield  {author} {\bibinfo {author} {\bibfnamefont {S.~M.}\ \bibnamefont
  {Young}}, \bibinfo {author} {\bibfnamefont {S.}~\bibnamefont {Zaheer}},
  \bibinfo {author} {\bibfnamefont {J.~C.}\ \bibnamefont {Teo}}, \bibinfo
  {author} {\bibfnamefont {C.~L.}\ \bibnamefont {Kane}}, \bibinfo {author}
  {\bibfnamefont {E.~J.}\ \bibnamefont {Mele}}, \ and\ \bibinfo {author}
  {\bibfnamefont {A.~M.}\ \bibnamefont {Rappe}},\ }\href@noop {} {\bibfield
  {journal} {\bibinfo  {journal} {Physical review letters},\ }\textbf {\bibinfo
  {volume} {108}},\ \bibinfo {pages} {140405} (\bibinfo {year}
  {2012})}\BibitemShut {NoStop}%
\bibitem [{\citenamefont {Schoop}\ \emph {et~al.}(2013)\citenamefont {Schoop},
  \citenamefont {MŸchler}, \citenamefont {Felser},\ and\ \citenamefont
  {Cava}}]{schoop2013lone}%
  \BibitemOpen
  \bibfield  {author} {\bibinfo {author} {\bibfnamefont {L.~M.}\ \bibnamefont
  {Schoop}}, \bibinfo {author} {\bibfnamefont {L.}~\bibnamefont {MŸchler}},
  \bibinfo {author} {\bibfnamefont {C.}~\bibnamefont {Felser}}, \ and\ \bibinfo
  {author} {\bibfnamefont {R.}~\bibnamefont {Cava}},\ }\href@noop {} {\bibfield
   {journal} {\bibinfo  {journal} {Inorganic chemistry},\ }\textbf {\bibinfo
  {volume} {52}},\ \bibinfo {pages} {5479} (\bibinfo {year}
  {2013})}\BibitemShut {NoStop}%
\bibitem [{\citenamefont {Steinberg}\ \emph {et~al.}(2014)\citenamefont
  {Steinberg}, \citenamefont {Young}, \citenamefont {Zaheer}, \citenamefont
  {Kane}, \citenamefont {Mele},\ and\ \citenamefont
  {Rappe}}]{steinberg2014bulk}%
  \BibitemOpen
  \bibfield  {author} {\bibinfo {author} {\bibfnamefont {J.~A.}\ \bibnamefont
  {Steinberg}}, \bibinfo {author} {\bibfnamefont {S.~M.}\ \bibnamefont
  {Young}}, \bibinfo {author} {\bibfnamefont {S.}~\bibnamefont {Zaheer}},
  \bibinfo {author} {\bibfnamefont {C.}~\bibnamefont {Kane}}, \bibinfo {author}
  {\bibfnamefont {E.}~\bibnamefont {Mele}}, \ and\ \bibinfo {author}
  {\bibfnamefont {A.~M.}\ \bibnamefont {Rappe}},\ }\href@noop {} {\bibfield
  {journal} {\bibinfo  {journal} {Physical review letters},\ }\textbf {\bibinfo
  {volume} {112}},\ \bibinfo {pages} {036403} (\bibinfo {year}
  {2014})}\BibitemShut {NoStop}%
\bibitem [{\citenamefont {Wan}\ \emph {et~al.}(2011)\citenamefont {Wan},
  \citenamefont {Turner}, \citenamefont {Vishwanath},\ and\ \citenamefont
  {Savrasov}}]{wan2011topological}%
  \BibitemOpen
  \bibfield  {author} {\bibinfo {author} {\bibfnamefont {X.}~\bibnamefont
  {Wan}}, \bibinfo {author} {\bibfnamefont {A.~M.}\ \bibnamefont {Turner}},
  \bibinfo {author} {\bibfnamefont {A.}~\bibnamefont {Vishwanath}}, \ and\
  \bibinfo {author} {\bibfnamefont {S.~Y.}\ \bibnamefont {Savrasov}},\
  }\href@noop {} {\bibfield  {journal} {\bibinfo  {journal} {Physical Review
  B},\ }\textbf {\bibinfo {volume} {83}},\ \bibinfo {pages} {205101} (\bibinfo
  {year} {2011})}\BibitemShut {NoStop}%
\bibitem [{\citenamefont {Yang}\ \emph {et~al.}(2011)\citenamefont {Yang},
  \citenamefont {Lu},\ and\ \citenamefont {Ran}}]{PhysRevB.84.075129}%
  \BibitemOpen
  \bibfield  {author} {\bibinfo {author} {\bibfnamefont {K.-Y.}\ \bibnamefont
  {Yang}}, \bibinfo {author} {\bibfnamefont {Y.-M.}\ \bibnamefont {Lu}}, \ and\
  \bibinfo {author} {\bibfnamefont {Y.}~\bibnamefont {Ran}},\ }\Doi
  {10.1103/PhysRevB.84.075129} {\bibfield  {journal} {\bibinfo  {journal}
  {Phys. Rev. B},\ }\textbf {\bibinfo {volume} {84}},\ \bibinfo {pages}
  {075129} (\bibinfo {year} {2011})}\BibitemShut {NoStop}%
\bibitem [{\citenamefont {Chevrel}\ \emph {et~al.}(1985)\citenamefont
  {Chevrel}, \citenamefont {Gougeon}, \citenamefont {Potel},\ and\
  \citenamefont {Sergent}}]{chevrel1985ternary}%
  \BibitemOpen
  \bibfield  {author} {\bibinfo {author} {\bibfnamefont {R.}~\bibnamefont
  {Chevrel}}, \bibinfo {author} {\bibfnamefont {P.}~\bibnamefont {Gougeon}},
  \bibinfo {author} {\bibfnamefont {M.}~\bibnamefont {Potel}}, \ and\ \bibinfo
  {author} {\bibfnamefont {M.}~\bibnamefont {Sergent}},\ }\href@noop {}
  {\bibfield  {journal} {\bibinfo  {journal} {Journal of Solid State
  Chemistry},\ }\textbf {\bibinfo {volume} {57}},\ \bibinfo {pages} {25}
  (\bibinfo {year} {1985})}\BibitemShut {NoStop}%
\bibitem [{\citenamefont {Struss}\ and\ \citenamefont
  {Corbett}(1969)}]{struss1969lower}%
  \BibitemOpen
  \bibfield  {author} {\bibinfo {author} {\bibfnamefont {A.~W.}\ \bibnamefont
  {Struss}}\ and\ \bibinfo {author} {\bibfnamefont {J.~D.}\ \bibnamefont
  {Corbett}},\ }\href@noop {} {\bibfield  {journal} {\bibinfo  {journal}
  {Inorganic Chemistry},\ }\textbf {\bibinfo {volume} {8}},\ \bibinfo {pages}
  {227} (\bibinfo {year} {1969})}\BibitemShut {NoStop}%
\bibitem [{\citenamefont {Lachgar}\ \emph {et~al.}(1990)\citenamefont
  {Lachgar}, \citenamefont {Dudis},\ and\ \citenamefont
  {Corbett}}]{lachgar1990revision}%
  \BibitemOpen
  \bibfield  {author} {\bibinfo {author} {\bibfnamefont {A.}~\bibnamefont
  {Lachgar}}, \bibinfo {author} {\bibfnamefont {D.~S.}\ \bibnamefont {Dudis}},
  \ and\ \bibinfo {author} {\bibfnamefont {J.~D.}\ \bibnamefont {Corbett}},\
  }\href@noop {} {\bibfield  {journal} {\bibinfo  {journal} {Inorganic
  Chemistry},\ }\textbf {\bibinfo {volume} {29}},\ \bibinfo {pages} {2242}
  (\bibinfo {year} {1990})}\BibitemShut {NoStop}%
\bibitem [{\citenamefont {Lepetit}\ \emph {et~al.}(1984)\citenamefont
  {Lepetit}, \citenamefont {Monceau}, \citenamefont {Potel}, \citenamefont
  {Gougeon},\ and\ \citenamefont {Sergent}}]{lepetit1984superconductivity}%
  \BibitemOpen
  \bibfield  {author} {\bibinfo {author} {\bibfnamefont {R.}~\bibnamefont
  {Lepetit}}, \bibinfo {author} {\bibfnamefont {P.}~\bibnamefont {Monceau}},
  \bibinfo {author} {\bibfnamefont {M.}~\bibnamefont {Potel}}, \bibinfo
  {author} {\bibfnamefont {P.}~\bibnamefont {Gougeon}}, \ and\ \bibinfo
  {author} {\bibfnamefont {M.}~\bibnamefont {Sergent}},\ }\href@noop {}
  {\bibfield  {journal} {\bibinfo  {journal} {Journal of low temperature
  physics},\ }\textbf {\bibinfo {volume} {56}},\ \bibinfo {pages} {219}
  (\bibinfo {year} {1984})}\BibitemShut {NoStop}%
\bibitem [{\citenamefont {Schoop}\ \emph {et~al.}(2014)\citenamefont {Schoop},
  \citenamefont {Hirschberger}, \citenamefont {Tao}, \citenamefont {Felser},
  \citenamefont {Ong},\ and\ \citenamefont {Cava}}]{schoop2014paramagnetic}%
  \BibitemOpen
  \bibfield  {author} {\bibinfo {author} {\bibfnamefont {L.}~\bibnamefont
  {Schoop}}, \bibinfo {author} {\bibfnamefont {M.}~\bibnamefont
  {Hirschberger}}, \bibinfo {author} {\bibfnamefont {J.}~\bibnamefont {Tao}},
  \bibinfo {author} {\bibfnamefont {C.}~\bibnamefont {Felser}}, \bibinfo
  {author} {\bibfnamefont {N.}~\bibnamefont {Ong}}, \ and\ \bibinfo {author}
  {\bibfnamefont {R.}~\bibnamefont {Cava}},\ }\href@noop {} {\bibfield
  {journal} {\bibinfo  {journal} {Physical Review B},\ }\textbf {\bibinfo
  {volume} {89}},\ \bibinfo {pages} {224417} (\bibinfo {year}
  {2014})}\BibitemShut {NoStop}%
\bibitem [{\citenamefont {Blaha}\ \emph {et~al.}(1990)\citenamefont {Blaha},
  \citenamefont {Schwarz}, \citenamefont {Sorantin},\ and\ \citenamefont
  {Trickey}}]{Blaha1990399}%
  \BibitemOpen
  \bibfield  {author} {\bibinfo {author} {\bibfnamefont {P.}~\bibnamefont
  {Blaha}}, \bibinfo {author} {\bibfnamefont {K.}~\bibnamefont {Schwarz}},
  \bibinfo {author} {\bibfnamefont {P.}~\bibnamefont {Sorantin}}, \ and\
  \bibinfo {author} {\bibfnamefont {S.}~\bibnamefont {Trickey}},\ }\Doi
  {10.1016/0010-4655(90)90187-6} {\bibfield  {journal} {\bibinfo  {journal}
  {Computer Physics Communications},\ }\textbf {\bibinfo {volume} {59}},\
  \bibinfo {pages} {399 } (\bibinfo {year} {1990})},\ ISSN \bibinfo {issn}
  {0010-4655}\BibitemShut {NoStop}%
\bibitem [{\citenamefont {Becke}\ and\ \citenamefont
  {Johnson}(2006)}]{becke:221101}%
  \BibitemOpen
  \bibfield  {author} {\bibinfo {author} {\bibfnamefont {A.~D.}\ \bibnamefont
  {Becke}}\ and\ \bibinfo {author} {\bibfnamefont {E.~R.}\ \bibnamefont
  {Johnson}},\ }\Doi {10.1063/1.2213970} {\bibfield  {journal} {\bibinfo
  {journal} {The Journal of Chemical Physics},\ }\textbf {\bibinfo {volume}
  {124}},\ \bibinfo {eid} {221101} (\bibinfo {year} {2006})}\BibitemShut
  {NoStop}%
\bibitem [{\citenamefont {Perdew}\ \emph {et~al.}(1996)\citenamefont {Perdew},
  \citenamefont {Burke},\ and\ \citenamefont
  {Ernzerhof}}]{perdew1996generalized}%
  \BibitemOpen
  \bibfield  {author} {\bibinfo {author} {\bibfnamefont {J.}~\bibnamefont
  {Perdew}}, \bibinfo {author} {\bibfnamefont {K.}~\bibnamefont {Burke}}, \
  and\ \bibinfo {author} {\bibfnamefont {M.}~\bibnamefont {Ernzerhof}},\
  }\href@noop {} {\bibfield  {journal} {\bibinfo  {journal} {Physical review
  letters},\ }\textbf {\bibinfo {volume} {77}},\ \bibinfo {pages} {3865}
  (\bibinfo {year} {1996})}\BibitemShut {NoStop}%
\bibitem [{\citenamefont {Aroyo}\ \emph {et~al.}(2014)\citenamefont {Aroyo},
  \citenamefont {Orobengoa}, \citenamefont {de~la Flor}, \citenamefont {Tasci},
  \citenamefont {Perez-Mato},\ and\ \citenamefont
  {Wondratschek}}]{aroyo2014brillouin}%
  \BibitemOpen
  \bibfield  {author} {\bibinfo {author} {\bibfnamefont {M.~I.}\ \bibnamefont
  {Aroyo}}, \bibinfo {author} {\bibfnamefont {D.}~\bibnamefont {Orobengoa}},
  \bibinfo {author} {\bibfnamefont {G.}~\bibnamefont {de~la Flor}}, \bibinfo
  {author} {\bibfnamefont {E.~S.}\ \bibnamefont {Tasci}}, \bibinfo {author}
  {\bibfnamefont {J.~M.}\ \bibnamefont {Perez-Mato}}, \ and\ \bibinfo {author}
  {\bibfnamefont {H.}~\bibnamefont {Wondratschek}},\ }\href@noop {} {\bibfield
  {journal} {\bibinfo  {journal} {Acta Crystallographica Section A: Foundations
  and Advances},\ }\textbf {\bibinfo {volume} {70}},\ \bibinfo {pages} {0}
  (\bibinfo {year} {2014})}\BibitemShut {NoStop}%
\end{thebibliography}%

\begin{figure}[htbp]

\begin{center}
\includegraphics[scale=0.3]{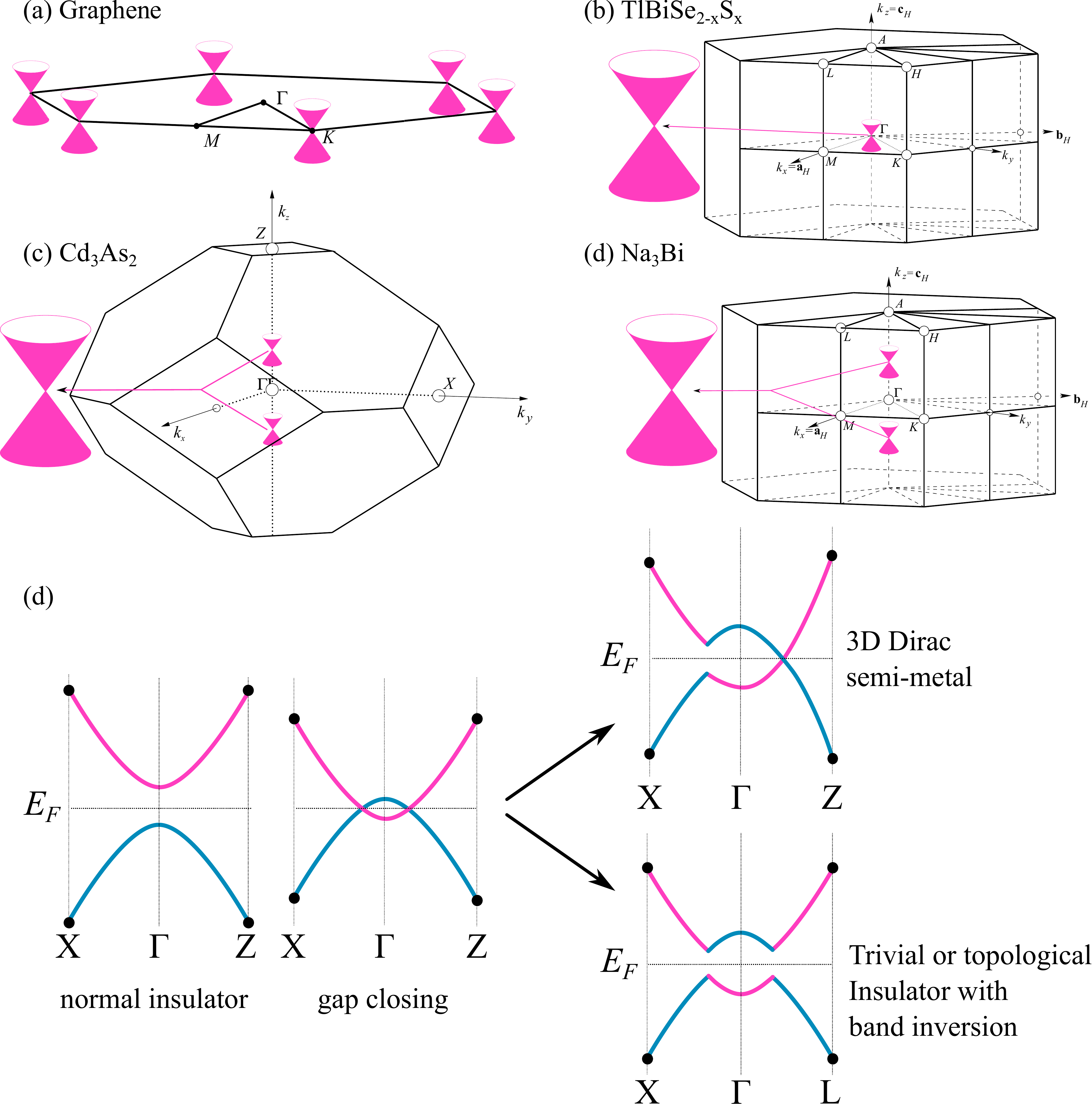}
\caption{(a-d) Schematic electronic structures showing the location of the 2D (graphene) or 3D Dirac cones in (a) graphene, (b)TlBiSe$_{2-x}$S$_x$, (c)Cd$_3$As$_2$, and (d)Na$_3$Bi. Brilloun zones redrawn from images at \cite{aroyo2014brillouin}(e) Schematic of a band gap closing, leading to either a Dirac semi-metal or band inverted insulator upon consideration of spin orbit coupling.}
\label{default}
\end{center}
\end{figure}

\begin{figure}[htbp]
\begin{center}
\includegraphics[scale=0.5]{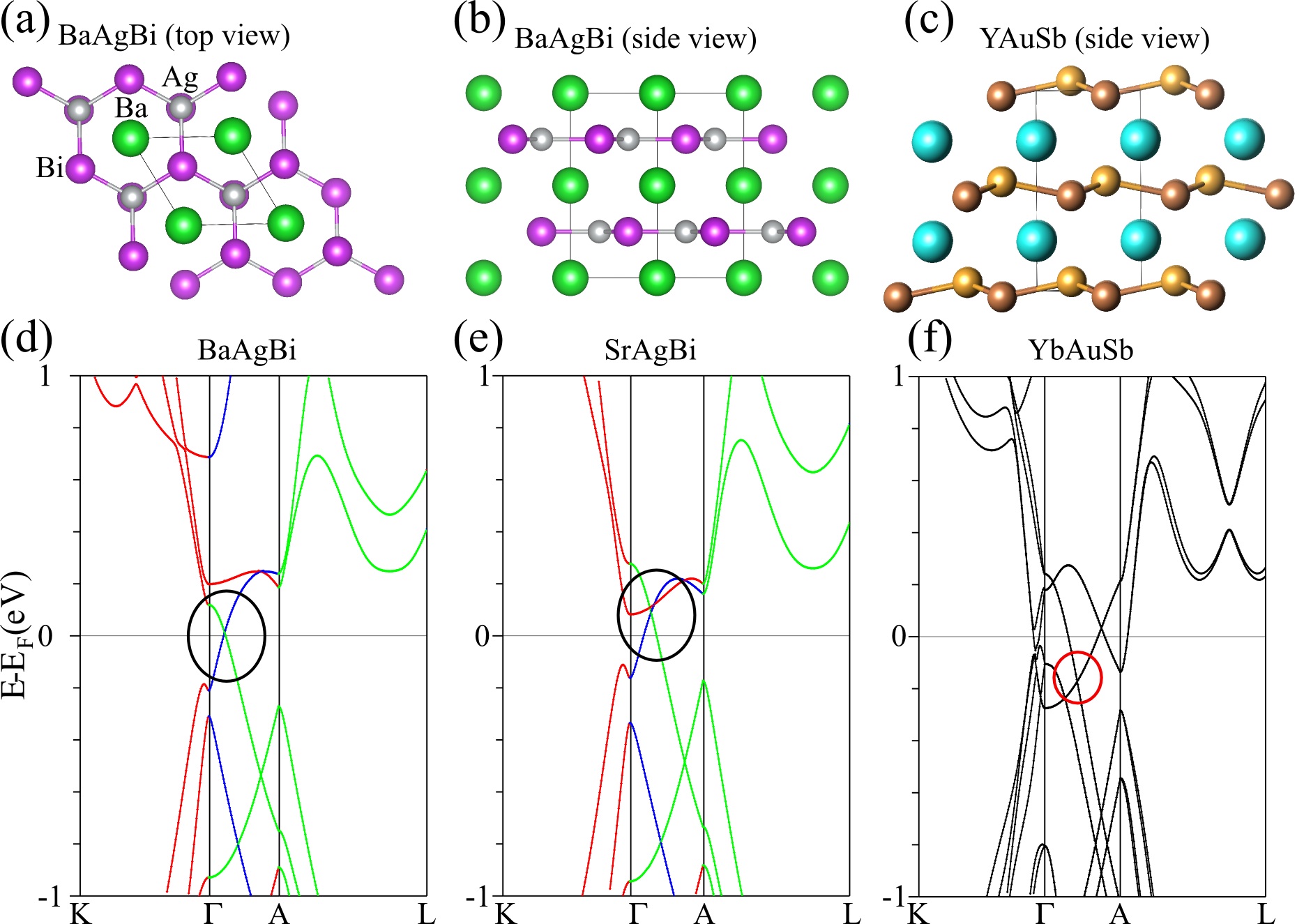}
\caption{Case I: 3D Dirac semi-metals based on charge balanced formulas (a) Top-down crystal structure of BaAgBi, representative of the entire ZrBeSi family. (b) Side view of BaAgBi, showing the 2 layer structure.(c) Side view of YbAuSb, showing the buckling of the honeycomb lattice (d) Electronic structure of BaAgBi; the 3D Dirac cone is circled in black. The different colored lines along $\Gamma$-A represent different irreducible representations under the C$6_v$ double group. (e) Electronic structure of the similar compound, SrAgBi. (f) Electronic structure of LiGaGe type buckled YbAuSb, calculated in the framework of LDA+U to account for the Yb core f electrons. (d-f) were calculated using the mbJ exchange functional which tends to have improved band gaps over PBE.}
\label{default}
\end{center}
\end{figure}

\begin{figure}[htbp]
\begin{center}
\includegraphics[scale=0.3]{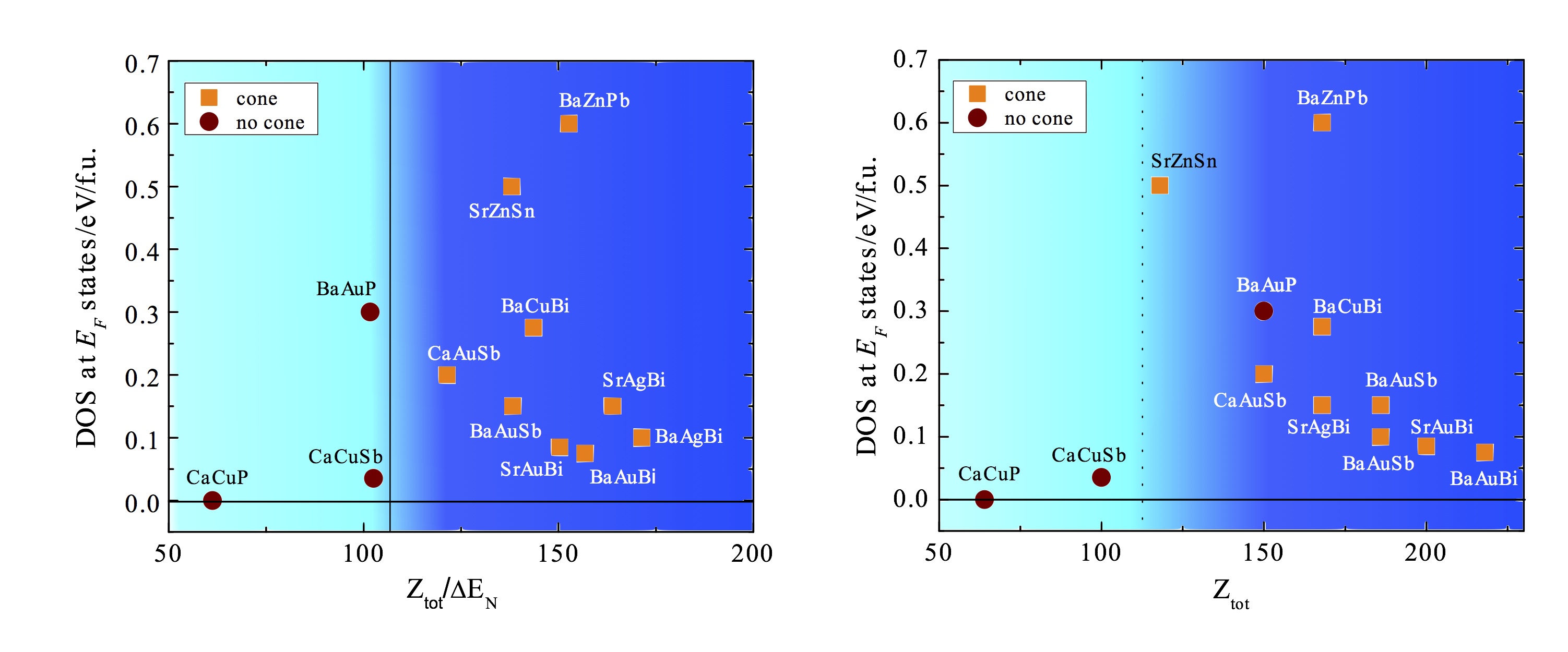}
\caption{Case I: 3D Dirac semi-metals based on charge balanced formulas (a) Electronic phase diagram of the ZrBeSi family, showing the density of states at E$_F$ as a function of the total Z divided by the electronegativity difference. Orange squares represent the compound having a Dirac cone, and red circles represent none. (b) Same as (a) but shown as a function of total Z only}
\label{default}
\end{center}
\end{figure}

\begin{figure}[htbp]
\begin{center}
\includegraphics[scale=0.5]{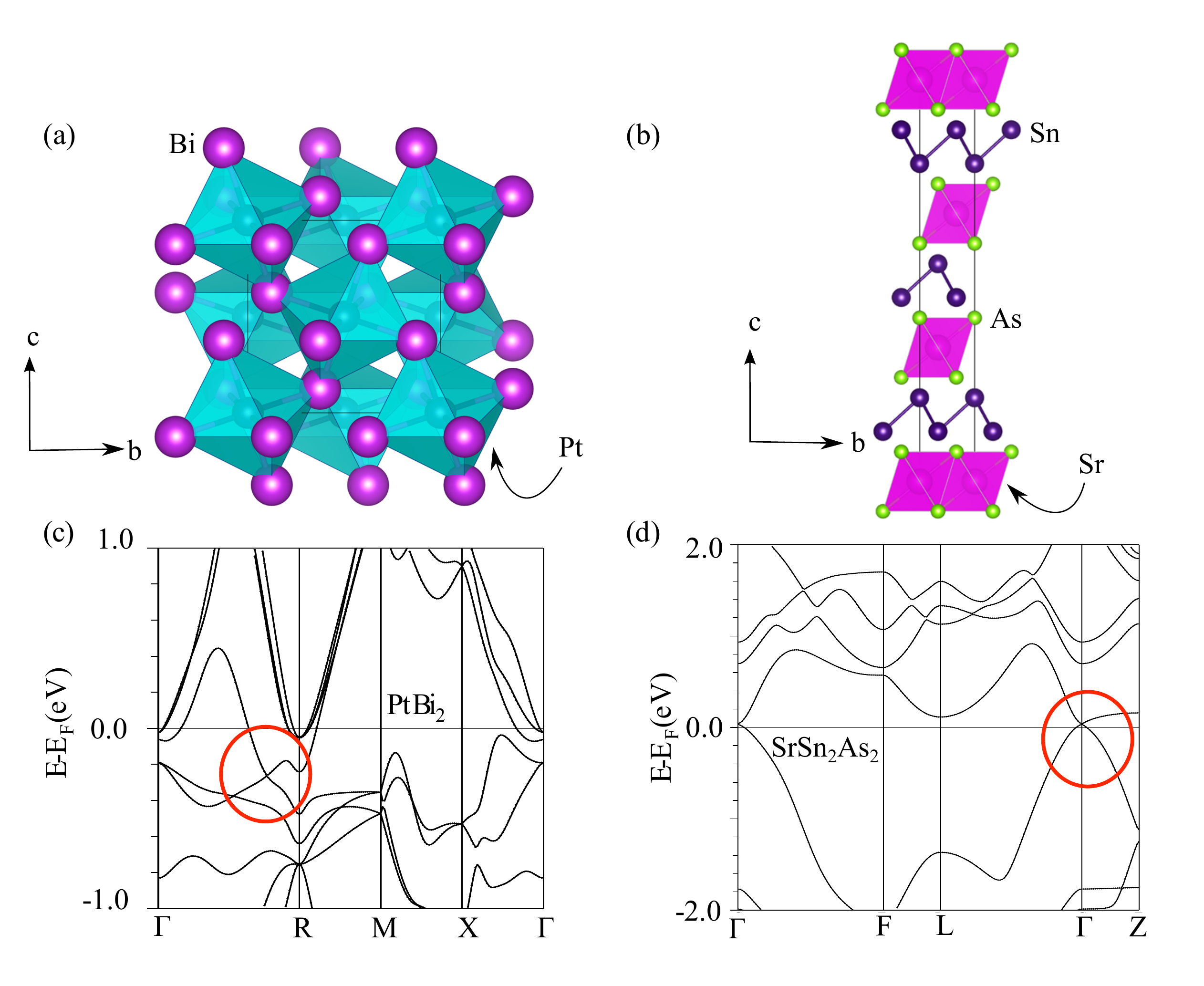}
\caption{Case I: 3D Dirac semi-metals based on charge balanced formulas.(a-b) Crystal structures of pyrite type PtBi$_2$ and tetradymite type SrSn$_2$As$_2$, left to right, respectively (c-d) Electronic structures of pyrite type PtBi$_2$ and tetradymite type SrSn$_2$As$_2$, left to right, respectively, both calculated with the mBJ functional.}
\label{default}
\end{center}
\end{figure}
 
\begin{figure}[htbp]
\begin{center}
\includegraphics[scale=0.5]{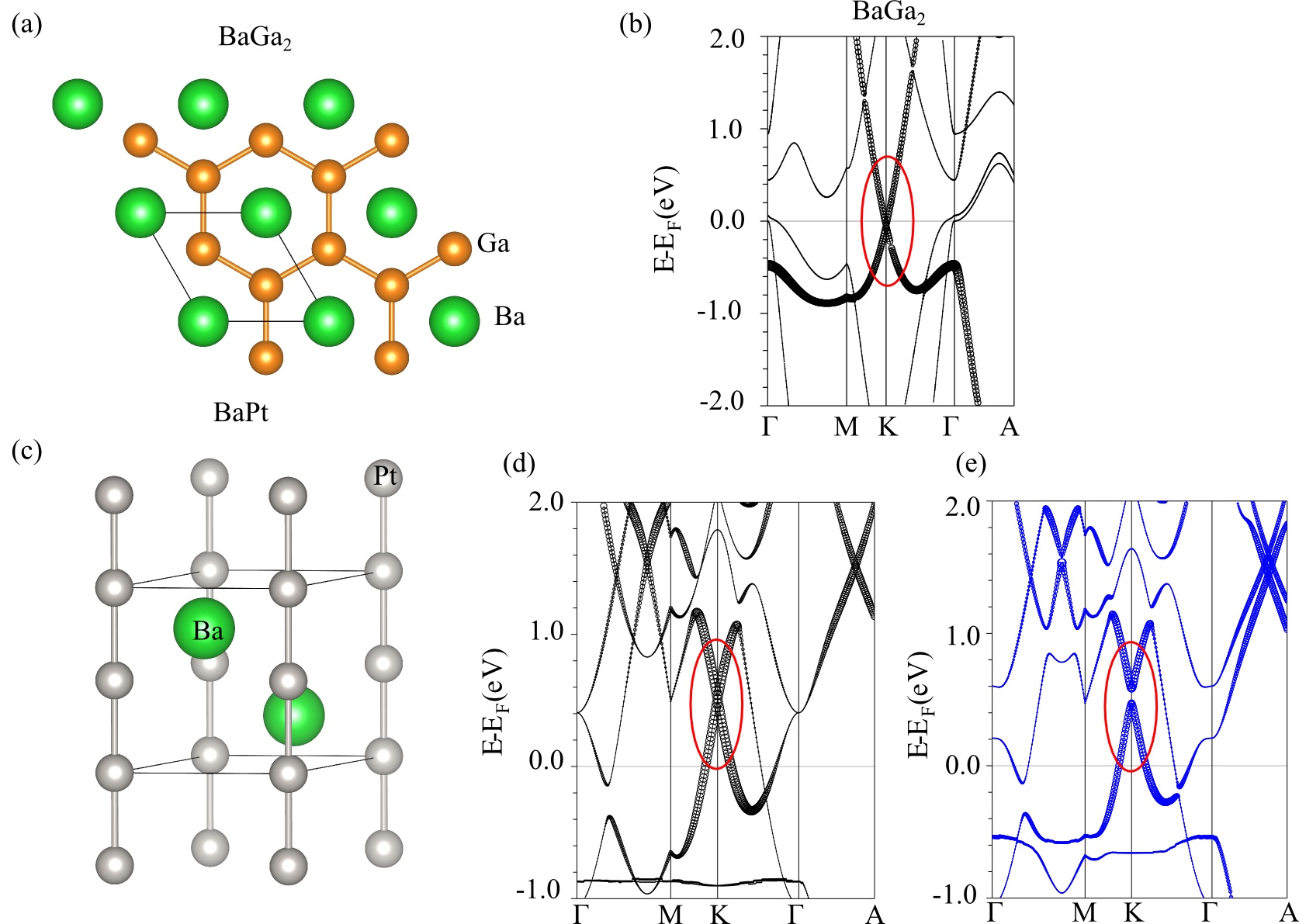}
\caption{Case II: 3D Dirac semi-metals in analogy to graphene (a) Top-down view of BaGa$_2$, showing the graphene like Ga sub lattice.  (b) Electronic structure of BaGa$_2$, with the Dirac cone circled in red. The fat bands show the contribution of the Ga p$_z$ orbitals. (c) Crystal structure of BaPt, with the Pt-Pt bonds drawn. (d) Electronic structure of BaPt, both without and with SOC (left to right, respectively). The Dirac cone is circled in red. The fat bands highlight the orbital contribution of the in plane d$_{xy}$ and d$_{x^2-y^2}$ orbitals of Pt. }
\label{default}
\end{center}
\end{figure}

\begin{figure}[htbp]
\begin{center}
\includegraphics[scale=0.5]{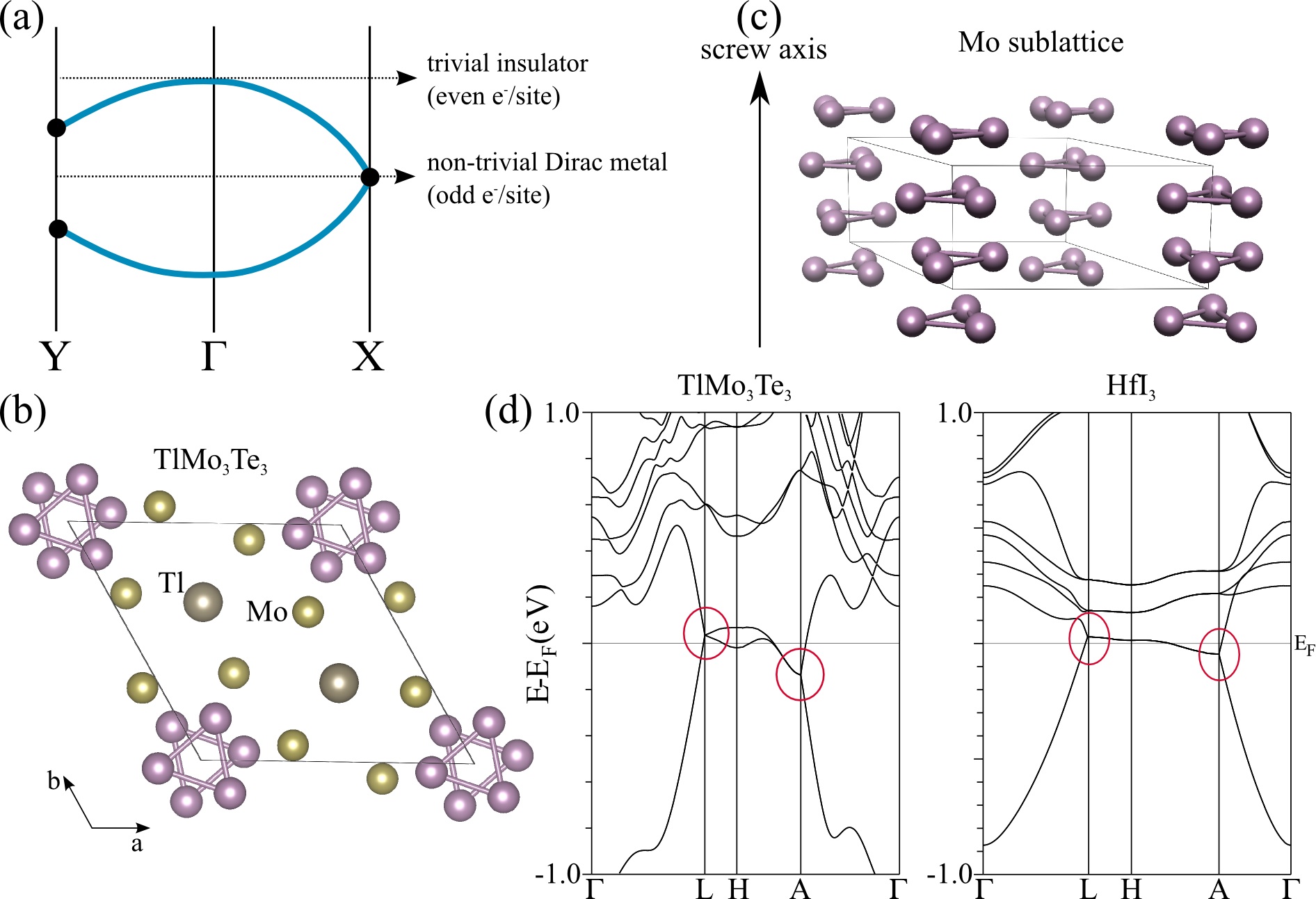}
\caption{Case IIIa: 3D Dirac semi-metals from non-symmorphic symmetry (a) Schematic of a band structure with non-symmorphic symmetry, and a sticking point at X. The electron counts for a normal insulator and a nontrivial metal are shown. (b) Top-down view of the crystal structure of TlMo$_3$Te$_3$, showing the Mo-Mo bonds. (c) A side view of the Mo$_3$ sub lattice of all AMo$_3$X$_3$ compounds. (d) Electronic structure of TlMo$_3$Te$_3$. The non-symmorphic sticking point that creates the anisotropic Dirac cone is circled in red. (e) Electronic structure of HfI$_3$. In this structure there are three non-symmorphic sticking points, creating a very quasi 1D Dirac cone.}
\label{default}
\end{center}
\end{figure}

\begin{figure}[htbp]
\begin{center}
\includegraphics[scale=0.5]{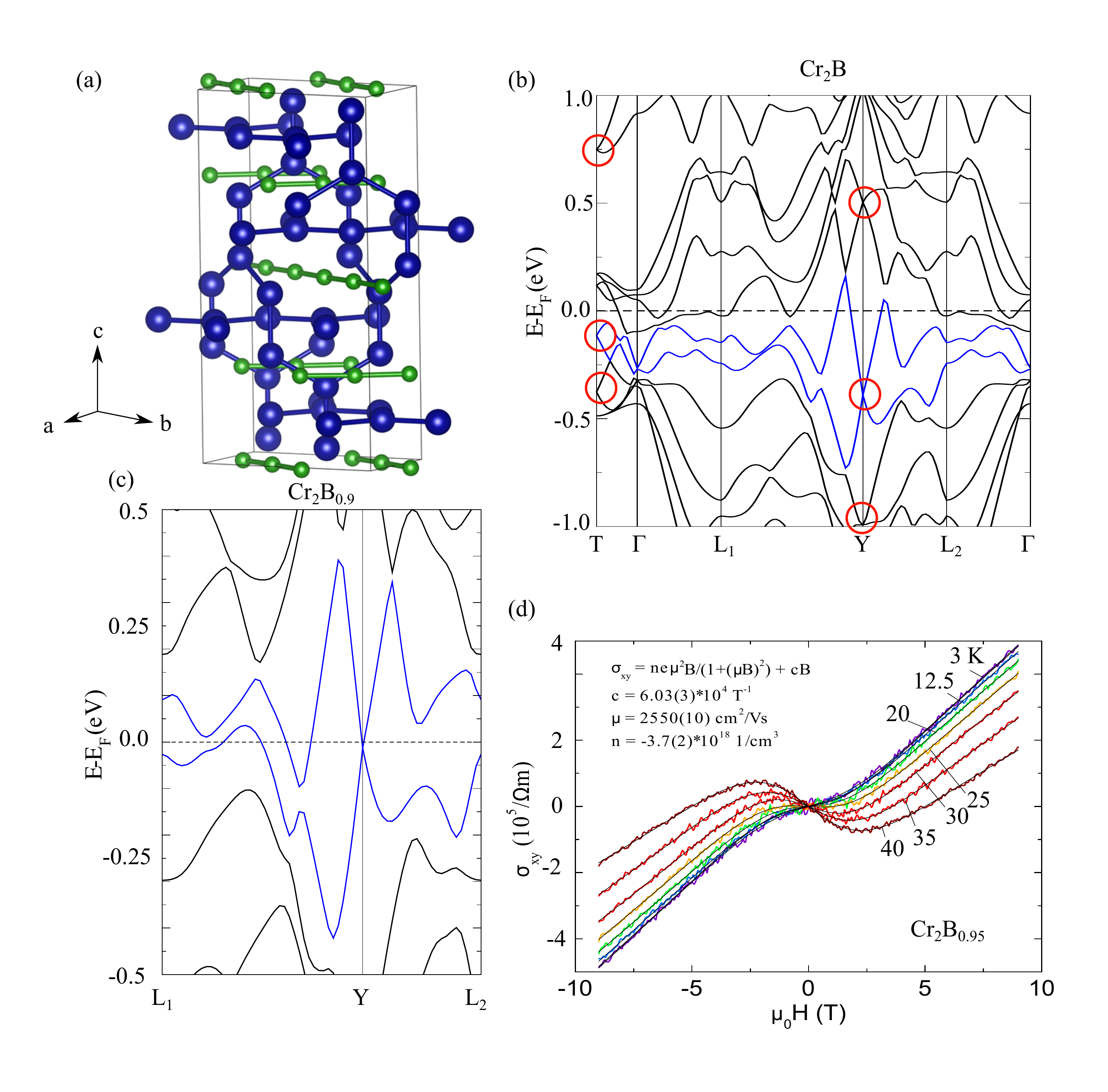}
\caption{Case IIIb: 3D Dirac cones in classical intermetallics with non-symmorphic symmetry (a) The crystal structure of Cr$_2$B. The interlocking honeycomb nets of Cr are shown by drawing the Cr-Cr bonds. (b) The electronic structure of stoichiometric Cr$_2$B. Some of the non-symmorphic sticking points are circled in red, and the bands making up the Dirac cone that is discussed are drawn in blue. (c) Close view of the electronic structure of Cr$_2$B$_{0.9}$, calculated using the virtual crystal approximation.(d) Hall effect measurements at various temperatures, with the fitted parameters for the 4K cure shown, to showcase the high mobility electron pocket.}
\label{default}
\end{center}
\end{figure}


\end{document}